\documentclass[
    preprint,
    5p,
    times,
    twocolumn,
    authoryear,
]{elsarticle}
\usepackage{adjustbox}
\usepackage{graphicx}
\usepackage{caption}
\usepackage{float}
\usepackage{placeins}
\usepackage{subcaption}
\usepackage[labelformat=simple]{caption}

\usepackage{amssymb,amsmath,amsthm}

\usepackage{booktabs}
\usepackage{ragged2e}
\usepackage{dcolumn}
\usepackage{makecell}
\usepackage{threeparttable}
\usepackage[switch]{lineno}
\usepackage{multirow}
\usepackage{hyperref}
\hypersetup{
    colorlinks=true,
    allcolors=blue,
    pdfauthor={Name}
}
\begin{document}

% FRONTMATTER
\begin{frontmatter}

\title{Last-layer committee machines for uncertainty estimations of benthic imagery}

\author[inst1]{H. Martin Gillis}
\author[inst1]{Isaac Xu}
\author[inst2,inst3]{Benjamin Misiuk}
\author[inst4]{Craig J. Brown}
\author[inst1]{Thomas Trappenberg\corref{cor1}}

\cortext[cor1]{Corresponding author. \\
E-mail address: \href{mailto:tt@cs.dal.ca}{tt@cs.dal.ca} (Thomas Trappenberg).}

\affiliation[inst1]{
    organization={Faculty of Computer Science, Dalhousie University},
    % addressline={}, 
    city={Halifax},
    % postcode={}, 
    state={Nova Scotia},
    country={Canada}
    }

\affiliation[inst2]{
    organization={Department of Earth Sciences, Memorial University of Newfoundland},
    % addressline={}, 
    city={St. John's},
    % postcode={}, 
    state={Newfoundland and Labrador},
    country={Canada}
    }

\affiliation[inst3]{
    organization={Department of Geography, Memorial University of Newfoundland},
    % addressline={}, 
    city={St. John's},
    % postcode={}, 
    state={Newfoundland and Labrador},
    country={Canada}
    }

\affiliation[inst4]{
    organization={Department of Oceanography, Dalhousie University},
    % addressline={}, 
    city={Halifax},
    % postcode={}, 
    state={Nova Scotia},
    country={Canada}
    }

\begin{abstract}  % 250 words or less
Automating the annotation of benthic imagery ({\it i.e.}, images of the seafloor and its associated organisms, habitats, and geological features) is critical for monitoring rapidly changing ocean ecosystems.
Deep learning approaches have succeeded in this purpose; however, consistent annotation remains challenging due to ambiguous seafloor images, potential inter-user annotation disagreements, and out-of-distribution samples.
Marine scientists implementing deep learning models often obtain predictions based on one-hot representations trained using a cross-entropy loss objective with softmax normalization, resulting with a single set of model parameters. 
While efficient, this approach may lead to overconfident predictions for context-challenging datasets, raising reliability concerns that present risks for downstream tasks such as benthic habitat mapping and marine spatial planning.
In this study, we investigated classification uncertainty as a tool to improve the labeling of benthic habitat imagery.
We developed a framework for two challenging sub-datasets of the recently publicly available BenthicNet dataset using Bayesian neural networks, Monte Carlo dropout inference sampling, and a proposed single last-layer committee machine.
This approach resulted with a $>95\text{\%}$ reduction of network parameters to obtain per-sample uncertainties while obtaining near-identical performance compared to computationally more expensive strategies such as Bayesian neural networks, Monte Carlo dropout, and deep ensembles.
The method proposed in this research provides a strategy for obtaining prioritized lists of uncertain samples for human-in-the-loop interventions to identify ambiguous, mislabeled, out-of-distribution, and/or difficult images for enhancing existing annotation tools for benthic mapping and other applications.
\end{abstract}

% \begin{graphicalabstract}
    % \includegraphics{TITLE}
% \end{graphicalabstract}

% \begin{highlights}
% \item Research highlight 1
% \item Research highlight 2
% \item Research highlight 3
% \end{highlights}


\end{frontmatter}

% INTRODUCTION
\section{Introduction}
\label{sect:introduction}
Our oceans require better tools for management and oversight, monitoring seafloor (benthic) habitats, and assessing environmental impact indicators~\citep{Winther::2020a}. 
Mapping benthic habitats has recently received much attention due to its importance in understanding changes in our oceans~\citep{Huang::2011b,Brown::2011a,Brown::2012a,Beijbom::2016a,Arosio::2023a,Misiuk::2024c}.
In part, this led to several large compilations of ocean floor imagery from research groups, government agencies, non-profit organizations, and other stakeholders from around the world~\citep{Brown::2012a,Katija::2022a,Humblot-Renaux::2024a,Misiuk::2024a,Lowe::2025a}.
For example, the recently publicly available BenthicNet~\citep{Lowe::2025a} consists of more than 11 million images, including \textit{ca.} 2.5 million annotations labeled according to the Collaborative and Automated Tools for Analysis of Marine
Imagery (CATAMI) hierarchal classification scheme~\citep{Althaus::2014a,Althaus::2015a}. 
It represents a diverse collection that can be used to develop and deploy deep learning networks to support automated mapping, understanding the seafloor environment, and policy development for sustainable oceans and environmental management.

Most contemporary supervised methods train Convolutional Neural Networks (CNNs) such as Residual Networks (ResNets)~\citep{He::2016a} or attention-based architecture such as Vision Transformers (ViTs)~\citep{Dosovitskiy::2021a} to predict a class label, given image data.
A common classification model represents class membership using a one-hot representation, which can be optimized using the cross-entropy loss function.
The softmax of the model outputs can be interpreted as the probability density function $p(y~|~\mathbf{x}; \mathbf{w})$, where $y$ are the class predictions, $\mathbf{x}$ are the image inputs for the network, and $\mathbf{w}$ are the trainable parameters of the model.
However, these networks are known to overestimate the confidence of their output predictions~\citep{Holm::2023a,Szegedy::2015a,Wei::2022a}.
In these cases, the normalized outputs of the model (\textit{i.e.}, softmax) often result with preferred predictions, even in cases where the predictions are incorrect.
This is particularly problematic for marine image data, which are notoriously difficult to annotate~\citep{Ovadia::2019a,Humblot-Renaux::2024a,Xu::2024a}. 
Oceanographer are therefore interested in tools that can provide them with prioritized per-sample reliability scores so that they can identify images that can be inspected manually.

Uncertainty in predictions can be derived from different sources~\citep{Abdar::2021a,Gawlikowski::2023a,Hullermeier::2021a}. For example, data (aleatoric) uncertainty is inherent to the data itself, such as measurement noise and/or mislabeling or annotation errors.
Collecting additional data would not reduce the uncertainty.
Another source is model (epistemic) uncertainty, which reflects the lack of knowledge (or data) for the model to describe.
Thus, in the limit of training with infinite data, model uncertainty can be reduced.
An example would be out-of-distribution images~\citep{Ovadia::2019a}, where a model attempts to make predictions for which it lacks representations (\textit{e.g.}, images from different geographic locations).
In this case, these images can also reduce performance and potential applications of deep learning networks for benthic mapping.
We therefore seek a strategy to identify uncertain or difficult images that can either be removed because of poor quality and/or submitted for subject matter expert interventions. 
In more general terms, we are interested in not only the most likely prediction for a single input image, but also how confident the network is for a difficult prediction.
In addition, we require a scalable approach for large benthic image datasets that are constantly evolving over time to provide reliable information for changing environmental conditions~\citep{Li::2023b}.
Given the recent advances for uncertainty estimations for deep neural networks~\citep{Abdar::2021a,Gawlikowski::2023a}, we seek an efficient strategy to identify a prioritized list of uncertain samples for Subject Matter Expert (SME) re-evaluations.
In this work, {\textit{we propose to use classification as an approach to identify uncertain samples by comparing common uncertainty evaluators such as Bayesian Model Averaging (BMA) and Monte Carlo Dropout (MCD) to a proposed single Last-Layer Committee Machine (LLCM)}}.

Our main motivation for using a single last-layer committee machine is to provide a modular and compute-efficient approach for rapid access to obtain per-sample uncertainties.
This approach goes beyond standard metrics (\textit{e.g.}, accuracy) that enables a human-in-the-loop strategy for re-evaluating difficult samples~\citep{Liu::2022a}.

An LLCM can provide this with a single forward pass of inputs during inference.
Specifically, we use only one fully-connected layer for each member of the committee machine.
This dramatically reduces the number of network parameters used to generate per-sample uncertainties.
This is in contrast to BMA and MCD approaches where the entire network is sampled multiple times (\textit{e.g.}, 100+) to obtain per-sample uncertainties.
For example, when using a feature extracting network with 100K parameters and a classification network with 1K parameters, sampling this network $100 \times$ (\textit{e.g.}, MCD) requires evaluating \textit{ca.}~10M parameters to obtain per-sample uncertainties. 
When using an LLCM with the same feature extracting network and a classification network defined with 100 last-layer committee members with 1K parameters each, the resulting network involves a total of 200K parameters (sampled only once).
This represents a \textit{ca.} 98\% reduction of network parameters to obtain per-sample uncertainties.

The gradients of each committee member of an LLCM module are independently backpropagated (\textit{i.e.}, no logit averaging) without the need for sophisticated loss functions or network configurations.
The requisite committee member diversity was obtained by randomly initializing network weights without applying additional techniques during training to promote diversity.
Random initialization and the stochastic nature of training were sufficient to provide a set of uncertain samples.
This was further investigated by comparing the coefficients of variation of learned committee members' network weights and singular values from singular value decompositions of network weights during and after training.

To generate prioritized lists of uncertain samples for different uncertainty evaluators, we applied increasing threshold values of confidence scores (\textit{i.e.}, maximum softmax probabilities).
The accuracies of the resulting subset of predictions (\textit{i.e.}, removal of difficult image predictions) were then plotted with respect to the fraction of samples remaining and confusion matrices were used to extract the corresponding remaining fractions of correct and incorrect model predictions to compute an efficiency metric.
This is an unbiased approach for removal of uncertain samples as it evaluates network predictions that were both correct and incorrect.
These uncertainty metric plots provided additional insights into the overall effectiveness of selecting different threshold values and comparing uncertainty evaluators.

The main contributions of our work can be summarized as the following:

\begin{itemize}
    \item We propose a single LLCM classifier providing an efficient and scalable approach to obtain per-sample uncertainties.
    \item We present uncertainty metric plots as a data analysis tool to provide information on network performance and selectivity based on selected threshold values.
    \item Lastly, based on our analyses of network weights, we provide support that random initializations of committee members is sufficient to obtain epistemic (model) uncertainty.
\end{itemize}

\noindent
To present our contributions, apart from this section, this paper is organized in the following sections: 1) Related work. In this section we discuss the challenges of annotating benthic images and techniques used to evaluate uncertainty estimates. 2) Methods. This section presents the framework for uncertainty evaluators, definition of the LLCM, loss calculation, and per-sample uncertainty estimations. In addition, this section includes descriptions for datasets, network configurations and calibration, and evaluation metrics. 3) Experiments. Here, we present our results to confirm network diversity of LLCM members and a comparison of uncertainty evaluators for a benchmark study and benthic images. 4) Conclusions. We provide concluding comments for the LLCM framework for improving uncertainty estimations for benthic imagery and other domain areas.

% RELATED
\section{Related work}
\label{sect:related}
Despite having access to high-quality and large-scale benthic images, these datasets can still contain difficult images for networks and human annotators~\citep{Brown::2012a,Misiuk::2024c,Humblot-Renaux::2024a}.
This can drastically impede the development of robust automated tools for benthic habitat mapping and decision-support for ocean management.
Combined with changing benthic habitats, network performance is continuously challenged with exposure to unseen images that can limit practical applications in real-world monitoring.
Many studies for benthic habitat mapping using neural networks focus on developing large datasets and applying transfer learning techniques to boost overall performance.

During their study of coral reefs in the Gulf of Eilat (Aqaba),~\cite{Raphael::2020a} acquired \textit{ca.} 5000K images with 11 different coral species.
The authors reported classification accuracies of 80\% when using pre-trained ResNet-50 networks and 90\% when using VGG-16~\citep{Simonyan::2014a} networks.
\cite{Yasir::2021a} also investigated the classification of coral reefs.
In this case, the authors incorporated two image enhancement strategies in addition to a pre-trained DenseNet-169 network to obtain an accuracy of 87\% for 9 classes.
Using a dataset of 70000 benthic images consisting of coral and substrate images (\textit{i.e.}, sand/mud, pebbles/gravel+cobbles, and rocks),~\cite{Jackett::2023a} obtained an accuracy of 98\% for a pre-trained ResNet-50 network.
The authors performed additional data pre-processing and applied several random augmentations such as, rotation, erasing, perspective, affine, and equalize during training.
As part of the BenthicNet dataset, the authors also provided their results for the Substrate (depth 2) and German Bank 2010 sub-datasets~\citep{Lowe::2025a}.
Using a pre-trained ResNet-50 network trained using a supervised (cross-entropy) or self-supervised (Barlow Twins) objective function, the authors obtained 88\% and 77\% accuracy for the Substrate (depth 2) and German Bank 2010 sub-datasets, respectively.

An important component for a robust annotating tool for benthic mapping is addressing potential distribution shifts.
This can result from a distal effect, where images from a geographic region are not represented in the current weights of the network.
\cite{Humblot-Renaux::2024a} (Denmark) recognized that the BenthicNet dataset was lacking images for their geographic region.
Subsequently,~\citeauthor{Humblot-Renaux::2024a} collected benthic images from video clips and created the JAMBO dataset (3290 images).
This dataset was then trained using several pre-trained networks to obtain cross-validation F1-scores of 92\%.
The authors also evaluated the inter-rater reliability for human annotations of the testing images ($N=6$), resulting with an uncertainty evaluation of 84\% for differentiating sand \textit{vs.} stone benthic images.
In addition to the distal effect on distribution shifts, there exists a temporal effect where geographic regions change over time and remains an open problem~\citep{Humblot-Renaux::2024a}.
Identifying a prioritized list of difficult or uncertain samples for human-in-the-loop interventions may offer an efficient strategy to address these distribution shifts.

Uncertainty estimates are often obtained using BMA of Bayesian Neural Networks (BNN)~\citep{Neal::1995a,Arbel::2023a,Jospin::2020a}, Bayesian approximating methods such as, Monte Carlo Dropout (MCD) inference sampling~\citep{Gal::2015a,Gal::2017a,Xie::2021a}, and deep ensembles~\citep{Lakshminarayanan::2017a,Pearce::2018a,Egele::2021a}.
However, these approaches have significant computational requirements as multiple networks are trained and evaluated (multiple times) during inference.
To address this issue, strategies such as SnapShot ensembles~\citep{Huang::2017a} and BatchEnsembles~\citep{Wen::2020c} were developed.
Another popular area of research to reduce computational requirements are ``last-head'' versions of deep ensemble networks whereby representation learning is decoupled from uncertainty estimations~\citep{Lee::2015a,Valdenegro-Toro::2023a,Harrison::2024a,Steger::2024a}.

\cite{Lee::2015a} introduced a TreeNet deep ensemble architecture as a last-head ensemble where each head consists of one or more convolutional blocks (\textit{i.e.}, a convolution and pooling layer) with the last block containing a fully-connected and softmax prediction layer. 
The authors highlighted that random initialization of shared network parameters can outperform full ensembles, dramatically reducing the computational requirements.
The authors proposed an ensemble aware loss function that seeks to promote network diversity \textit{via} multiple choice learning.
In contrast to the proposed LLCM, a single fully-connected layer is used for each committee member (\textit{i.e.}, no convolution or pooling layers), where the module consists of 100 committee members as opposed to the 5 heads reported by~\citeauthor{Lee::2015a}.
In the case of our last-layer network, diversity was also derived from the random initialization of parameters using the cross-entropy loss function for training.

\cite{Valdenegro-Toro::2023a} followed-up on the \citeauthor{Lee::2015a} study, proposing deep sub-ensembles as an approach to obtain fast estimations of uncertainties.
The authors used a similar last-head strategy, defining a single trunk network for representation learning and multiple task networks for uncertainty estimations.
The main difference of deep sub-ensembles is the trunk network is initially trained and then the weights are fixed before training with multiple task networks.
Compared to the LLCM, deep sub-ensembles approach uses multiple network layers for each head and involves additional pre-training steps to obtain uncertainties.
The authors report using up to 15 heads for their deep sub-ensembles, compared to the 100 committee machine members we used for this study, which is made possible by extending a single LLCM.
LLCMs provide an end-to-end approach where uncertainty estimations are obtained in a single forward-pass of inputs.

Perhaps closest to our work in terms of last-layer \textit{vs.}~last-head strategies are two recent works of~\citeauthor{Harrison::2024a}~\citep{Harrison::2024a} and~\citeauthor{Steger::2024a}~\citep{Steger::2024a} in 2024.
\citeauthor{Harrison::2024a} proposed using a Variational Bayesian Last-Layer Network (VBLL), while~\citeauthor{Steger::2024a} introduced Particle-Optimization Variational Inference (POVI) for a last-layer network trained in function space.
These techniques can be used to separate representation learning and uncertainty estimations, which would be applicable for pre-trained networks.

While these last-head approaches (\textit{i.e.}, TreeNets and deep sub-ensembles) reduced computational requirements, they come with additional steps to promote network diversity and/or specialized training procedures.
Whereas the last-layer Bayesian techniques (\textit{i.e.}, VBLL, and POVI) are often difficult to configure for large datasets, such as BenthicNet.
The LLCM further simplifies these techniques by using a single last-layer module and random initialization of network weights.
In this study, we compare the LLCM to MCD inference sampling and fully Bayesian networks for a benchmark experiment and last-layer Bayesian networks for the more challenging benthic imagery.
We show that LLCMs provide comparable uncertainty estimations and obtain results in an efficient single forward-pass of inputs.

% METHODS
\section{Methods}
\label{sect:methods}
In this section, we describe uncertainty evaluators and our proposed LLCM.
We provide experimental details relating to data collection and processing, network configurations, hyperparameters, and network calibrations.
We then demonstrate this approach to identify uncertain samples using different uncertainty evaluators: BMA, MCD, and our proposed LLCMs.
Using the challenging benthic images from the publicly available high-resolution BenthicNet dataset~\citep{Misiuk::2024a,Lowe::2025a}, we were able to obtain uncertain samples for subject matter expert interventions.

\subsection{Uncertainty evaluators}
An approach for quantifying model uncertainty are Bayesian (or equivalent approximation) models, where a distribution of models are derived from learned parameters.
Therefore, in terms of Bayesian analysis, we assess not only one model, but all possible models given the training data $\mathcal{D}$.
The posterior predictive distribution (\autoref{eq:bayesian_integral}) is computed by marginalizing all possible parameters (\textit{i.e.}, by integration), often using a standard Gaussian prior $\mathcal{N}(\mathbf{w};0, \sigma{^2}\mathbf{I})$~\citep{kendall_what_2017,blundell_weight_2015}).

\begin{align}
    p(y|\mathbf{x}, \mathcal{D}) 
    &= \int_{\mathbf{w}} 
    \overbrace{
    p(y | \mathbf{x};\mathbf{\mathbf{w}})
    }^{likelihood}
    \underbrace{
    p(\mathbf{\mathbf{w}})
    }_{prior}
    d\mathbf{w} \label{eq:bayesian_integral} \\
    &\approx \frac{1}{M} \sum_{m=1}^{M} p(y|\mathbf{x}; \mathbf{w}^m)
    \label{eq:bayesian_approximation}
\end{align}

\noindent
The objective function is defined through the likelihood term 
$p(y | \mathbf{x};\mathbf{\mathbf{w}})$, whose logarithm corresponds to the standard training loss (\textit{e.g.}, negative log-likelihood or cross-entropy), combined with a regularization term derived from the prior.
However, for most cases this approach is computationally intractable.
To address this challenge, Bayesian approximation methods were developed such as Monte Carlo dropout inference sampling~\citep{Gal::2015a} and deep ensembles~\citep{Lakshminarayanan::2017a}, including the last-layer strategies discussed in~\autoref{sect:related}.
This is achieved by averaging network predictions (\autoref{eq:bayesian_approximation}) where $M$ represents the number of models and individual probability distributions are calculated from output logits using network weights $\mathbf{w}^m$ and the softmax function.
%
% \begin{table}[H]
\begin{table}[!tb]
    \centering
    \begin{adjustbox}{width=0.48\textwidth}
    \begin{threeparttable}
        \caption{Symbols and notations.}
        \begin{tabular}{ p{2.5cm} p{6cm} }
            \toprule
            \textbf{Symbol} &
            \textbf{Description}
            \\
            \midrule
            $\mathcal{D}$ & 
            \makecell[l]{Dataset} 
            \\
            $\mathbf{x}$  & 
            \makecell[l]{Input data} 
            \\
            $y$ & 
            \makecell[l]{Output data} 
            \\
            $\mathbf{w}$ & 
            \makecell[l]{Weights of the model} 
            \\
            $\mathbf{w}^m$ & 
            \makecell[l]{Weights of the $m$\textsuperscript{th} model in the ensemble} 
            \\  
            $p(y|\mathbf{x}, \mathcal{D})$ & 
            \makecell[l]{Posterior predictive distribution} 
            \\
            $\mathbf{x} \rightarrow f(\mathbf{x}; \mathbf{w}^f)$ & 
            \makecell[l]{Feature extraction function} 
            \\
            $g^{m}(f; \mathbf{w}^{g^m})$ & 
            \makecell[l]{Uncertainty estimation function for the \\  $m$\textsuperscript{th} committee member} 
            \\
            $\mathcal{L}$ & 
            \makecell[l]{Total loss} 
            \\
            $\ell^m(y, g^{m}(f; \mathbf{w}^{g^m}))$ & 
            \makecell[l]{Loss for the $m$\textsuperscript{th} committee member} 
            \\
            $M$ & 
            \makecell[l]{Number of models or committee members} 
            \\
            $\mathbf{w^{g^m}}$ & 
            \makecell[l]{Weights of the $m$\textsuperscript{th} model for uncertainty \\ estimation} 
            \\
            $\mu$ & 
            \makecell[l]{Mean value for image normalization \\ (\textit{e.g.}, for BenthicNet)} 
            \\
            $\sigma$ & 
            \makecell[l]{Standard deviation for image normalization \\ (\textit{e.g.}, for BenthicNet)} 
            \\
            $N$ & 
            \makecell[l]{Number of samples} 
            \\
            $p(y_n|\mathbf{x}_n)$ & 
            \makecell[l]{Predicted probability for the $n$\textsuperscript{th} sample} 
            \\
            $\mathbb{I}(y_n = 1)$  & 
            \makecell[l]{Indicator function for the true label $y_n = 1$} 
            \\
            $B_m$ & 
            \makecell[l]{Set of samples in the $m$\textsuperscript{th} bin \\ (for ECE calculation)} 
            \\ 
            $A_k(B_m)$ & 
            \makecell[l]{Accuracy for class $k$ in the $m$\textsuperscript{th} bin \\ (for ECE calculation)}  
            \\
            $C_k(B_m)$ & 
            \makecell[l]{Mean predicted confidence for class $k$ in the \\ $m$\textsuperscript{th} bin (for ECE calculation)} 
            \\
            $K$ & 
            \makecell[l]{Number of classes} 
            \\
            $|B_m|$ & 
            \makecell[l]{Number of samples in the $m$\textsuperscript{th} bin} 
            \\
            ${\text{CV}}\tnote{1}$ & 
            \makecell[l]{Coefficient of variation calculated as $\sigma / \mu$} 
            \\
            $\sigma_{\text{CV}_M}$ & 
            \makecell[l]{Standard deviation of the CV ($\mathbf{w}$) \\ across $M$ committee members} 
            \\
            $Z_m$ & 
            \makecell[l]{Output weight matrix for committee \\ member $m$ of dimension $m \times n$}
            \\
            $U_m$ & 
            \makecell[l]{Left singular vectors of dimension $m \times m$} 
            \\
            $\Sigma_m$ & 
            \makecell[l]{Singular values matrix of dimension $m \times n$}
            \\
            $V_m^T$ & 
            \makecell[l]{Right singular vectors of dimension $n \times n$}
            \\
            ${\text{SV}^{i}_{m}}$ & 
            \makecell[l]{The $i$\textsuperscript{th} singular value from the singular \\ value matrix $\mathbf{\Sigma}{_m}$ of the $m$\textsuperscript{th} committee \\ member}
            \\
            $\text{CV}_{\text{SV}^{i}_{M}}$ & 
            \makecell[l]{Coefficient of variation of the \\ respective ${\text{SV}^{i}_{m}}$ across $M$ \\ committee members}
            \\
            $\text{CV}_{\|\text{SV}\|_{F}}$ & 
            \makecell[l]{Coefficient of variation of the Frobenius \\ norm of singular values across $M$ \\ committee members} 
            \\
            \midrule
            \bottomrule
        \end{tabular}
        $^1$ 
        % CV denotes the coefficient of variation $\sigma / \mu$ used as a normalized measure of dispersion to quantify diversity.
        CV provides a scale-independent measure of dispersion, making it an effective measure for network diversity since it quantifies relative variability across committee members.
        A higher CV indicates greater variability (diversity) across models.
        Details for specific CV-based diversity measures are provided in the text.
        \label{tab:symbols}
    \end{threeparttable}
    \end{adjustbox}
\end{table}

% \FloatBarrier

\begin{figure}[!tb]
    \centering
    \includegraphics[width=0.48\textwidth]{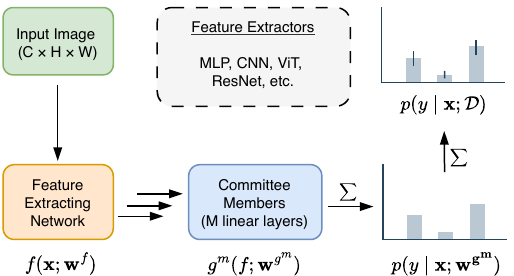}
    \caption{Last-layer committee machine architecture.
    An input image $\mathbf{x} \in \mathbb{R}^{C\times H\times W}$ is mapped by a feature-extracting network $f(\mathbf{x};\mathbf{w}^{f})$ (\textit{e.g.}, MLP, CNN, ViT) to a shared representation.
    This representation is fed to a last-layer committee machine defined with $M$ independently initialized linear classifiers $g^{m}(f;\mathbf{w}^{g^{m}})$, each producing class probabilities $p(y\mid\mathbf{x};\mathbf{w}^{g^{m}})$.
    During training, all members see identical features but optimize separate losses $\ell^{m}$.
    At inference, member softmax outputs are averaged to obtain the posterior predictive distribution $p(y\mid\mathbf{x},\mathcal{D})$ and associated uncertainties.}
  \label{fig:last-layer-committee}
\end{figure}

% LLCM
\paragraph{Last-layer committee machines}
A last-layer committee machine (or ensemble) is a form of an ensemble classifier used to boost model performance and resulting predictions by averaging multiple models, often used with classification and regression trees such as random forests.
Rather than using the full network as an ensemble, we used a list of $M$ linear layers as a modular component for the network architecture (\autoref{fig:last-layer-committee}~and~\autoref{eq:llcm}).
\begin{equation}
    \mathbf{x} \rightarrow 
    f(\mathbf{x}; \mathbf{w}^f) \xrightarrow{m} 
    g^{m}(f; \mathbf{w}^{g^m}) \rightarrow 
    p(y|\mathbf{x}; \mathbf{w^{g^m}})
    \rightarrow
    p(y|\mathbf{x}, \mathcal{D})
    \label{eq:llcm}
\end{equation}

\noindent
This approach separates the network into a two distinct networks for representation learning $f(\mathbf{x}; \mathbf{w}^f)$ of inputs $\mathbf{x}$ and uncertainty estimation $g^{m}(f; \mathbf{w}^{g^m})$.
In this case, representations of input images $\mathbf{x}$ are created (\textit{e.g.}, ResNet-50) and passed to multiple uncertainty networks $g^m$, where each uncertainty network (\textit{i.e.}, committee member) is randomly initialized with different sets weights.
Rather than backpropagating mean logit outputs of committee members, gradients of the computed loss of each committee member $\ell^m(y, g^{m}(f; \mathbf{w}^{g^m}))$ are independently backpropagated as loss $\mathcal{L}$ (\autoref{eq:llcm_loss}).

\begin{equation}
    \mathcal{L} =  \ell^1(y, g^{1}(f; \mathbf{w}^{g^1})) + 
    \cdots + \ell^m(y, g^{m}(f; \mathbf{w}^{g^m}))
    \label{eq:llcm_loss}
\end{equation}

\noindent
For every forward-pass during training, each committee member $m$ receives identical feature representations $f(\mathbf{x}; \mathbf{w}^f)$; therefore, network diversity or loss exploration during backpropagation is dependent on the initialization of each member.
If committee machine members are identically initialized, they will all have the same weights during and after training, where the loss landscape exploration will be identical for each member.
However, if they are non-identically initialized, the loss landscape exploration will be unique for a given committee member~\citep{Gawlikowski::2023a}.
During inference, individual probability distributions $p(y|\mathbf{x}; \mathbf{w^{g^m}})$ are calculated from output logits from using network weights $\mathbf{w^{g^m}}$ and the softmax function, which are then averaged over $M$ committee members to obtain the posterior predictive distribution as per-sample uncertainties (\autoref{eq:llcm_softmax}).

\begin{equation}
    p(y|\mathbf{x}, \mathcal{D}) \approx 
    \frac{1}{M} \sum_{m=1}^M 
    \frac{e^{g^{m}(f; \mathbf{w}^{g^m})}}{\sum e^{g^{m}(f; \mathbf{w}^{g^m})}}
    \label{eq:llcm_softmax}
\end{equation}

\noindent
These simple modifications results with a network architecture that can provide epistemic uncertainty for both the dataset and single samples at inference.
It drastically reduces computational and memory intensive requirements and can be parallelized and/or scaled to multiple devices.
In addition, LLCMs can be used with different feature extracting networks to easily separate feature representations from uncertainty estimations.

\subsection{Data collection and processing}
% MNIST
\paragraph{MNIST}
The MNIST dataset used for the benchmark study was obtained from the PyTorch dataset library and used without any further processing.
Training, validation, and testing datasets were created using 50000 samples for training and 10000 samples each for validation and testing datasets.
Class weights were also computed for the training dataset.

% GB
\paragraph{German Bank 2010}
The German Bank dataset consists of 3181 samples of five classes that describe the seafloor environment such as: silt/mud, silt with bedforms, reef, glacial till, and sand with bedforms.
Training and testing datasets were created by using the provided partition labels available in the annotation file~\citep{Misiuk::2024a}.
This resulted with a training dataset consisting of 2681 samples of silt/mud (23\%), silt with bedforms (7\%), reef (20\%), glacial till (27\%), and sand with bedforms (23\%).
The testing dataset contains 500 samples of silt/mud (42\%), silt with bedforms (6\%), reef (20\%), glacial till (7\%), and sand with bedforms (25\%).
The validation dataset was created by performing an 80:20 split of the training dataset, with label partitioning to ensure a balanced representation of class labels in both datasets.
Class weights were then computed for the resulting training dataset.
Images were resized to $224 \times 224$ and normalized using the BenthicNet corresponding means ($\mu = [0.359, 0.413, 0.386]$) and standard deviations ($\sigma = [0.219, 0.215, 0.209]$).

% S2
\paragraph{Substrate (depth 2)}
The Substrate dataset consists of 57149 samples of five classes that describe the seafloor environment such as: boulders, cobbles, rocks, pebbles/gravel, and sand/mud.
As was done for the German Bank dataset, the training and testing datasets were created using the provided partition labels available in the annotation file~\citep{Misiuk::2024a}.
In this case, the training dataset consists of 43430 samples of boulders (5\%), cobbles (2\%), rocks (13\%), pebbles/gravel (6\%), and sand/mud (74\%).
Whereas, the testing dataset contains 13719 samples of boulders (3\%), cobbles (2\%), rocks (17\%), pebbles/gravel (5\%), and sand/mud (73\%).
The validation dataset was created as described above.
Class weights were computed for the training dataset and all images were resized and normalized as described above.

\begin{table*}[!tb]
    \centering
    \begin{adjustbox}{width=\textwidth}
    \begin{threeparttable}
        \caption{Network configurations and hyperparameters.}
        \begin{tabular}{ l l l l l l l l l l}
            \toprule
            \textbf{Dataset} 
            & \textbf{Network}\tnote{1--6}
            & \textbf{Epochs} 
            & \textbf{Optimizer} 
            & \textbf{Learning Rate} 
            & \textbf{Scheduler} 
            & \textbf{Criterion} 
            & \textbf{Weight Decay} 
            & \textbf{Label Smoothing} 
            \\
            \midrule
            \multirow{3}{*}{MNIST}
            & BNN
            & 25
            & ClippedAdam
            & $1.0 \times 10^{-3}$
            & None
            & ELBO
            & None
            & None
            \\
            & CNN
            & 100
            & Adam
            & $1.0 \times 10^{-5}$
            & None
            & CrossEntropyLoss
            & 0.0
            & 0.0 or 0.1
            \\
            & LLCM
            & 100 
            & Adam
            & $1.0 \times 10^{-5}$
            & None
            & CrossEntropyLoss
            & 0.0 
            & 0.0 or 0.1
            \\
            \midrule
            \multirow{3}{*}{\makecell{German\\Bank}}
            & ResNet-BNN
            & 200
            & ClippedAdam (1.0)
            & $1.75 \times 10^{-3}$
            & None
            & ELBO
            & None
            & None
            \\
            & ResNet-CNN
            & 50
            & Adam
            & $3.0 \times 10^{-6}$
            & OneCycleLR
            & CrossEntropyLoss
            & $1.0 \times 10^{-5}$
            & 0.0 or 0.1
            \\
            & ResNet-LLCM
            & 25
            & Adam
            & $3.0 \times 10^{-6}$
            & OneCycleLR
            & CrossEntropyLoss
            & $1.0 \times 10^{-5}$
            & 0.0 or 0.1
            \\
            \midrule
            \multirow{3}{*}{Substrate}
            & ResNet-BNN
            & 15
            & ClippedAdam
            & $1.0 \times 10^{-3}$
            & None
            & ELBO
            & None
            & None
            \\
            & ResNet-CNN
            & 25
            & Adam
            & $1.0 \times 10^{-5}$
            & OneCycleLR
            & CrossEntropyLoss
            & $1.0 \times 10^{-3}$
            & 0.0 or 0.1
            \\
            & ResNet-LLCM
            & 25
            & Adam
            & $3.0 \times 10^{-6}$
            & OneCycleLR
            & CrossEntropyLoss
            & $1.0 \times 10^{-5}$
            & 0.0 or 0.1
            \\
            \midrule
            \bottomrule
        \end{tabular}
        $^1$ BNN: $\texttt{[Conv}(128,5,5)\!-\!\texttt{ReLU}\!-\!\texttt{MaxPool}(2,2)]_2 \to \texttt{Linear}(2048,10)$
        $^2$ CNN: $\texttt{[Conv}(128,5,5)\!-\!\texttt{ReLU}\!-\!\texttt{MaxPool}(2,2)\!-\!\texttt{Dropout}(p)]_2 \to \texttt{Linear}(2048,10)$
        $^3$ LLCM: $\texttt{[Conv}(128,5,5)\!-\!\texttt{ReLU}\!-\!\texttt{MaxPool}(2,2)\!-\!\texttt{Dropout}(p)]_2 \to [\texttt{Linear}(2048,10)]_{100}$
        $^4$ ResNet-BNN: $\texttt{ResNet(BottleNecks)} \to [\texttt{Linear}(2048,2048)\!-\!\texttt{BatchNorm}\!-\!\texttt{ReLU}] \to \texttt{Linear}(2028,5)$
        $^5$ ResNet-CNN: $\texttt{ResNet(BottleNecks)} \to [\texttt{FC}(2048,2048)\!-\!\texttt{BatchNorm}\!-\!\texttt{ReLU}\!-\!\texttt{Dropout}(p)] \to \texttt{Linear}(2048,5)$
        $^6$ ResNet-LLCM: $\texttt{ResNet(BottleNecks)} \to [\texttt{Linear}(2048,2048)\!-\!\texttt{BatchNorm}\!-\!\texttt{ReLU}\!-\!\texttt{Dropout}(p)] \to [\texttt{Linear}(2048,5)]_{100}$
        \label{tab:network_configurations}
    \end{threeparttable}
    \end{adjustbox}
\end{table*}

% \FloatBarrier

% MNIST
\subsection{Network configurations, hyperparameters, and calibrations}

\paragraph{Bayesian neural networks}
For the MNIST dataset a Bayesian neural network (BNN) was created by converting a baseline CNN using utilities available from the Pyro framework~\citep{Bingham::2019a} with Gaussian weight priors initialized to a mean of 0 and a standard deviation of 1.
The baseline network consisted of a feature extraction module with two blocks of convolutions, ReLU activations, and a maxpooling layer.
Both blocks used convolution with 128 output channels, a kernel size of 5, and a maxpooling layer with a kernel and stride size of 2.
The second block was subsequently flattened before being passed to a classifier which consisted of a linear layer (2048 input and 10 output channels).
Training (\textit{i.e.}, stochastic variational inference) was performed over 25 epochs using default values of the Pyro ClippedAdam optimizer with a learning rate of $1.0 \times 10^{-3}$ and the evidence lower bound (ELBO) loss objective function using the Categorical distribution represented by logits and target labels (\autoref{tab:network_configurations}).
In the case of the BenthicNet datasets, last-layer Bayesian networks~\citep{Harrison::2024a} were created using a BenthicNet pre-trained ResNet-50 model~\citep{Xu::2024a}.
The fully-connected component was modified with a linear layer (2048 input and output channels, respectively) followed by batch normalization, and ReLU activation.
A classifier was added which consisted of a linear layer (2048 input and 5 output channels) which was converted to a Bayesian layer using utilities from the Pyro framework.
The German Bank 2010 dataset (batch size 128) was trained over 200 epochs using the Pyro ClippedAdam optimizer with a learning rate of $1.75 \times 10^{-3}$ and default values, except for gradient clip normalization which was set to 1.0.
The Substrate (depth 2) dataset (batch size 128) was trained over 15 epochs with a learning rate of $1.0 \times 10^{-3}$ with all other settings set to their default values (\autoref{tab:network_configurations}).

% CNN
\paragraph{Convolutional neural networks}
Using the above baseline CNN for the MNIST dataset, a dropout layer was added after each pooling layer with a dropout rate of 0.1.
The dataset (batch size 128) was trained over $100$ epochs using the Adam optimizer with a learning rate of $1.0 \times 10^{-5}$ and the cross-entropy loss objective function with/without class weights and label smoothing~\citep{Szegedy::2015a} or with/without logit normalization~\citep{Wei::2022a} (\autoref{tab:network_configurations}).
For the BenthicNet datasets, a BenthicNet pre-trained ResNet-50 model~\citep{Xu::2024a} was used with the addition of a dropout layer ($p=0.01$) after each ReLu activation of the BottleNeck block of the ResNet-50 architecture.
The fully-connected component was modified with a linear layer (2048 input and output channels, respectively) followed by batch normalization, ReLU activation, and a dropout layer (p = 0.01).
A classifier was added which consisted of a linear layer (2048 input and 5 output channels).
The German Bank 2010 dataset (batch size 128) was trained over 50 epochs using the Adam optimizer with a learning rate of $3.0 \times 10^{-6}$, weight decay of $1.0 \times 10^{-5}$, and the cross-entropy loss function with/without class weights, logit normalization, or label smoothing.
The OneCycleLR scheduler~\citep{Smith::2017a,PyTorch::2025a} was applied during training using an initial and final learning rate factor of 0.1, where the rate was increased over the first 10\% of the total number of epochs. 
This was then repeated for the Substrate (depth 2) dataset over 25 epochs (\autoref{tab:network_configurations}).

% LLCM
\paragraph{Last-layer committee machines}
For the MNIST dataset, a dropout layer was added after each pooling layer with a dropout rate of 0.1 and a list of committee members of size $100$ as the classifier for the previously described baseline CNN.
Training was done with a batch size of 128 over 100 epochs using the Adam optimizer with a learning rate of $1.0 \times 10^{-5}$ and the cross-entropy loss objective function (for each $M$ classifier), with/without class weights and label smoothing or with/without logit normalization.
The total loss, mean probabilities, and corresponding standard deviations were computed as previously described (\autoref{sect:methods} and \autoref{tab:network_configurations}).
For the ResNet-CNN models described above, the classifier was converted to a list of 100 classifiers with 5 output channels for the BenthicNet datasets.
Training was preformed over 25 epochs with a learning rate of $3.0 \times 10^{-6}$, weight decay of $1.0 \times 10^{-5}$, using the OneCycleLR scheduler as previously described.
For both datasets, training was done using the cross-entropy loss objective function (for each $M$ classifier), with/without class weights and label smoothing or with/without logit normalization.
The total loss was then scaled by the number of committee members $M$ (\textit{i.e.}, a scaling factor), before being backpropagated during training.
This scaling factor can be considered an additional hyperparameter for training the network.
During inference, mean probabilities and corresponding standard deviations were computed by averaging the normalized logit outputs from $M$ committee members using the softmax function  (\autoref{tab:network_configurations}).

\paragraph{Network calibration}
Model calibration was performed by finding the optimal temperature (\textit{i.e.}, scaling of logits) that minimizes the cross-entropy loss on the validation dataset. 
This was achieved using a pre-trained model and Bayesian optimization with Gaussian Process (GP) regression~\citep{Louppe::2016a}. 
The optimization explored a temperature search space of 0.01--10.0, using a log-uniform prior which was configured to perform 50 evaluations of the cross-entropy objective function.
For LLCM models, the temperature for each committee member was optimized separately.
After finding the optimal temperature(s), the previously trained model was used to perform inference on the testing dataset, scaling the logits with the optimized calibrated temperature(s).
Network calibration was then evaluated based on commonly used metrics such as: Negative Log-likelihood (NLL), Brier Score (BS), Expected Calibration Error (ECE) and Reliability Diagrams~\citep{Guo::2017a}
as described int he next section.

\paragraph{Network evaluation and metrics}
\label{par:eval_metrics}
For applied applications, probabilities for predictions would ideally reflect observed accuracies, where predictions indicate how likely they are to be correct~\citep{Guo::2017a}.
For example, given network predictions with 80\% probability, we expect the fraction correct would be 80\% (perfectly calibrated).
In general, Calibration Error (CE) can be measured as the difference of average confidence and accuracy.
A negative calibration error indicate the model is under-confident and a positive calibration error indicate the model is over-confident.
This is particularly important for applications involving high-risk decision-making.
To evaluate the quality of model calibrations, several common metrics are used such as NLL, BS, ECE, and reliability diagrams.

\noindent
The NLL is defined as the negative logarithm of the likelihood function, which represents the probability of the observed data given the model parameters (\autoref{eq:nll}).

\begin{equation}    
    \text{NLL} = -\frac{1}{N} \sum_{n=1}^{N} \log p(y_n|\mathbf{x}_n)
    \label{eq:nll}
\end{equation}

\noindent
The BS is calculated as the mean squared error of $p(y|\mathbf{x})$ for the true predicted class and the corresponding one-hot encoded representations (\autoref{eq:bs}),

\begin{equation}    
    \text{BS} = \frac{1}{N} \sum_{n=1}^{N} (p(y_n|\mathbf{x}_n) - \mathbb{I}(y_n = 1))^2
    \label{eq:bs}
\end{equation}

\noindent
where
$N$ is the total number of samples; 
$p(y_n|\mathbf{x}_n)$ is the predicted probability for the positive class for the $n^{th}$ sample; and 
$\mathbb{I}(y_n = 1)$ is the indicator function, which is 1 if the true label $y_n$ is 1 and 0 otherwise.
The BS provides a metrics to score how well the dataset is calibrated with a value of 0 being perfectly calibrated.

\noindent
For a more discretized evaluation of the quality of model calibration, calibration errors can be calculated by partitioning confidence values into bins using a multiclass ECE as defined by (\autoref{eq:ece}),

\begin{equation}
   \text{ECE} = \sum_{m=1}^{M} \frac{|B_m|}{n} \sum_{k=1}^{K} \left| A_k(B_m) - C_k(B_m) \right| 
   \label{eq:ece}
\end{equation}

\noindent
where
$M$ is the number of bins;
$B_m$ is the set of samples whose predicted probabilities for the positive class are in the $m^{th}$ bin;
$|B_m|$ is the number of samples in the $m^{th}$ bin;
$n$ is the total number of samples;
$K$ is the number of classes;
$A_k(B_m)$ is the accuracy for class $k$ in the $m^{th}$ bin; and
$C_k(B_m)$ is the mean probability (confidence) for class $k$ in the $m^{th}$ bin.

Both the BS and ECE provides a scalar value that can be used to assess the quality of model calibrations; however, they do not distinguish whether a model is over- and/or under-confident.
While the CE metric does provides an overall assessment (\textit{i.e.}, under- \textit{vs.} over-confident), it is insufficient for ranges of confidence values.
In this case, a reliability diagram provides additional visual insights by comparing the mean predicted confidence in increasing discretized bins with the corresponding accuracy for a given bin.
A well-calibrated model will produce a positive linear relationship, indicating that predicted probabilities align with correct model predictions. 
Deviations from this relationship provides insights whether the model is under- (values above the line) and/or over-confident (values below the line) in its predictions.
We provide reliability diagrams with a histogram of confidence scores for each bin to assess the importance based on the total number of samples (normalized).
Once an evaluation of model calibration is obtained, a calibration method can be applied as a post-processing step (if necessary).
This is typically done with a validation dataset where a hyperparameter is adjusted to optimize model calibration.
For classification tasks, temperature scaling~\citep{Gawlikowski::2023a,Guo::2017a} during softmax normalization is an efficient method for model calibration.
After training, a validation dataset is used for optimization (\textit{e.g.}, Bayesian optimization) to find the ideal temperature, followed by a re-evaluation of model calibration.
In addition to the above network evaluations, we also report accuracy for the each of the the different approaches for BNN, CNN, LLCM, and different hyperparameters (\textit{e.g.}, label smoothing and logit normalization) for addressing over-confident network predictions.

% EXPERIMENTS
\section{Experiments}
\label{sect:experiments}

For the subsequent sections, we provide empirical support for random initialization using LLCM, which avoids requiring specialized techniques to promote network diversity.
We then compare our LLCM with two commonly used approaches to obtain uncertainty estimations: 1) Bayesian model averaging of BNNs; and 2) Monte Carlo dropout inference sampling of CNNs.
\textit{It is important to note that our focus here is not necessarily to identify a better performing model; rather, to find an equivalent performing model with reduced compute requirements and complexities.}
In the following sections, we examine network diversity, a benchmark study dataset and the results for the more challenging German Bank 2010 and Substrate (depth 2) datasets.

\subsection{Network diversity}
\label{sect:network_diversity}

While developing the last-layer committee machines for benthic imagery, we wanted to investigate network diversity across each committee member.
In particular, we wanted to ensure sufficient diversity persists at the end of training that would support using random initialization of weights.
We explored network diversity using statistical analyses of the coefficient of variations (CV) of committee member weights and corresponding singular value decompositions.
As others have noted, random initialization is sufficient for deep ensemble last-head and Bayesian last-layer approaches~\citep{Gawlikowski::2023a,Lee::2015a,Steger::2024a}.
Building on these concepts, LLCM ensemble-based uncertainty estimation restricts diversity to the final classification layer. 
Each committee member functions as a deterministic network with independent initialization, allowing the members to converge to local minima while reducing the computational cost associated with full ensemble networks.

\paragraph{Committee member weights}
Our initial approach to investigate network diversity used a trained LLCM-10 ($p=0.0)$ network and the MNIST dataset. 
After training, each committee member classifier weights were flattened and the coefficient of variations were computed.
Committee members with similar CV values would equate to being similar.
In this case, we expect the standard deviations of CV across committee members to be zero.
To confirm this hypothesis, we prepared a network where all committee members weights were initialized with ones.
\autoref{tab:cv_mnist} shows the results of using different hyperparameters during training, such as class weighting, logit normalization, and label smoothing. 
Although not required for this analysis, calibration metrics are provided for comparison.
The last entry of \autoref{tab:cv_mnist} shows the result where all committee machine members had weights initialized with ones.
This resulted with a $\sigma_{\text{CV}_M}$ of 0.0, which indicates that all members are identical.

The above approach provides support for network diversity and is a simple and convenient method; however, learned structural information may be lost as a result of exclusively using learned classifier weights values.
In addition, a large increase in network diversity ($\sigma_{\text{CV}_M}$) was observed when using logit normalization.
While this is expected given that logit normalization operates by factoring the magnitude of weight vectors, it does highlight potential numerical stability issues when calculating CV values (\textit{i.e.}, $\text{CV}=\sigma / \mu$).

\begin{table}[!tb]
    \centering
    \begin{adjustbox}{width=0.48\textwidth}
    \begin{threeparttable}
        \caption{Network diversity ($\sigma_{\text{CV}_M}$) and model performance of a LLCM-10 and the MNIST dataset.}
        \begin{tabular}{ l l l l l l l D{.}{.}{-1} }
            \toprule
            % headings
            \textbf{Model}\tnote{1}
            & \textbf{Accuracy / F1$\uparrow$}
            & \textbf{NLL$\downarrow$}
            & \textbf{BS$\downarrow$}
            & \textbf{ECE$\downarrow$}
            & \textbf{$\sigma_{\text{CV}_M}$}
            \\
            \midrule
            % entry: logs/multicnn10/01
            $-$/$-$/$-$
            & 0.991 / 0.991
            & 0.014
            & 0.003
            & 0.004
            & 9.537
            \\
            % entry: logs/multicnn10/02
            $-$/$+$/$-$
            & 0.990 / 0.989
            & 0.250
            & 0.072
            & 0.181
            & 906.146
            \\
            % entry: logs/multicnn10/03
            $+$/$+$/$-$
            & 0.989 / 0.989
            & 0.257
            & 0.074
            & 0.185
            & 8372.234
            \\
            % entry: logs/multicnn10/04
            $+$/$-$/$-$
            & 0.991 / 0.991
            & 0.014
            & 0.003
            & 0.003
            & 9.414
            \\
            % entry: logs/multicnn10/05
            $+$/$-$/$+$
            & 0.992 / 0.992
            & 0.141
            & 0.026
            & 0.115
            & 25.502
            \\
            % entry: logs/multicnn10/06
            $-$/$-$/$+$
            & 0.992 / 0.992
            & 0.141
            & 0.026
            & 0.115
            & 23.713
            \\
            \midrule
            % entry: logs/multicnn10/--
            $-$/$-$/$-$\tnote{2}
            & 0.991 / 0.991
            & 0.013
            & 0.003
            & 0.001
            & \textbf{0.0}
            \\
            \midrule
            \bottomrule
        \end{tabular}
        $^1$ Models defined based on training hyperparameters using class weights / logit normalization / label smoothing (amount of smoothing, 0.1). These hyperparameters are either applied (denoted by $+$) or omitted (denoted by $-$).
        $^2$ All committee machine members weights were initialized with ones. F1-scores reported as macro-averaged.
        \label{tab:cv_mnist}
    \end{threeparttable}
\end{adjustbox}
\end{table}

% \FloatBarrier

\paragraph{Committee member singular value decompositions}
To address the limitations of weight dispersion, we further examined structural diversity among committee members using singular value decomposition (SVD).
This section aims to provide a matrix-based view of how differently each member learns internal representations. 
While the following equations describe the formal computation of singular values and their derived measures, the central idea is that by decomposing the weight matrix of each member into its underlying components, we can quantify how much unique structure each member has learned.
By comparing the CV of respective singular values (SV) across members allows us to measure diversity in a way that is less sensitive to raw weight magnitude and more reflective of functional differences between models.

SVD is a linear algebra technique used to decompose a matrix into three corresponding matrices involving a rotation, re-scaling, and then another rotation, which is defined as
$\mathbf{Z}_m = \mathbf{U}_m \mathbf{\Sigma}_m \mathbf{V^{T}_m}$,
where $\mathbf{Z}_m$ is an output matrix for committee member $m$ of dimension $m \times n$; 
left singular vectors $\mathbf{U}{_m} \in \mathbb{R}^{m \times m}$;
singular values $\mathbf{\Sigma}{_m} \in \mathbb{R}^{m \times n}$; 
and right singular vectors $\mathbf{V^{T}_m} \in \mathbb{R}^{n \times n}$.
The diagonal of the $\mathbf{\Sigma}{_m}$ represents the singular values (\textit{i.e.}, unique values) of the $m^{th}$ committee member of size $\min (m, n)$.
A single scalar metric can be computed as the CV of the Frobenius norm of the SV for all committee members ($\text{CV}_{\|\text{SV}\|_F}$).
The Frobenius norm of the SV represents the overall magnitude (or energy).
This metric measures the variation across all committee members, which would be 0 if members are identical.
This strategy offers to maintain structural information and is less sensitive to outliers and numerical stability issues as compared to the above method.
By comparing respective CV and Frobenius norm values of the singular values across each committee member, an evaluation of network diversity can be quantitatively measured.

\begin{figure}[!tb]
    \centering
    \begin{subfigure}{0.24\textwidth}
        \includegraphics[width=\textwidth]{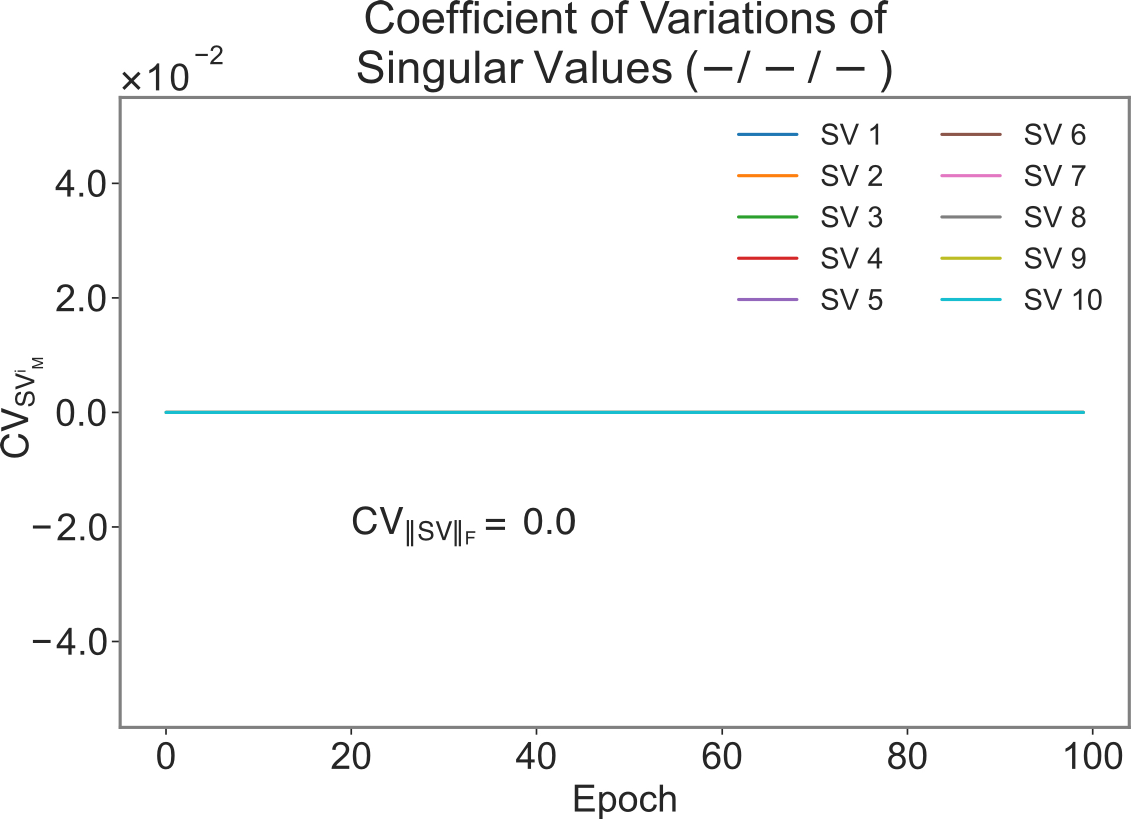}
        \caption{}
        \label{fig:svd_100}
    \end{subfigure}
    \hspace{-0.15cm}
    \vspace{0.1cm}
    \begin{subfigure}{0.24\textwidth}
        \includegraphics[width=\textwidth]{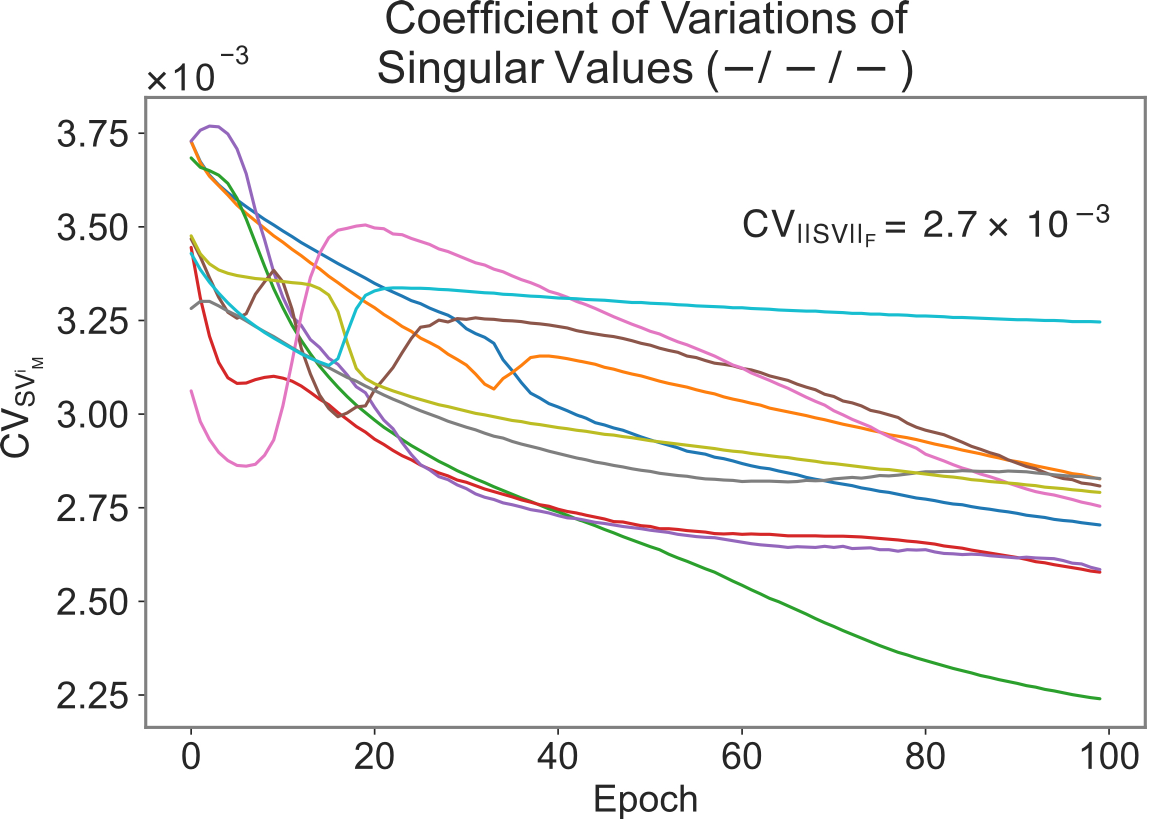}
        \caption{}
        \label{fig:svd_random_100}
    \end{subfigure}
    \begin{subfigure}{0.24\textwidth}
        \includegraphics[width=\textwidth]{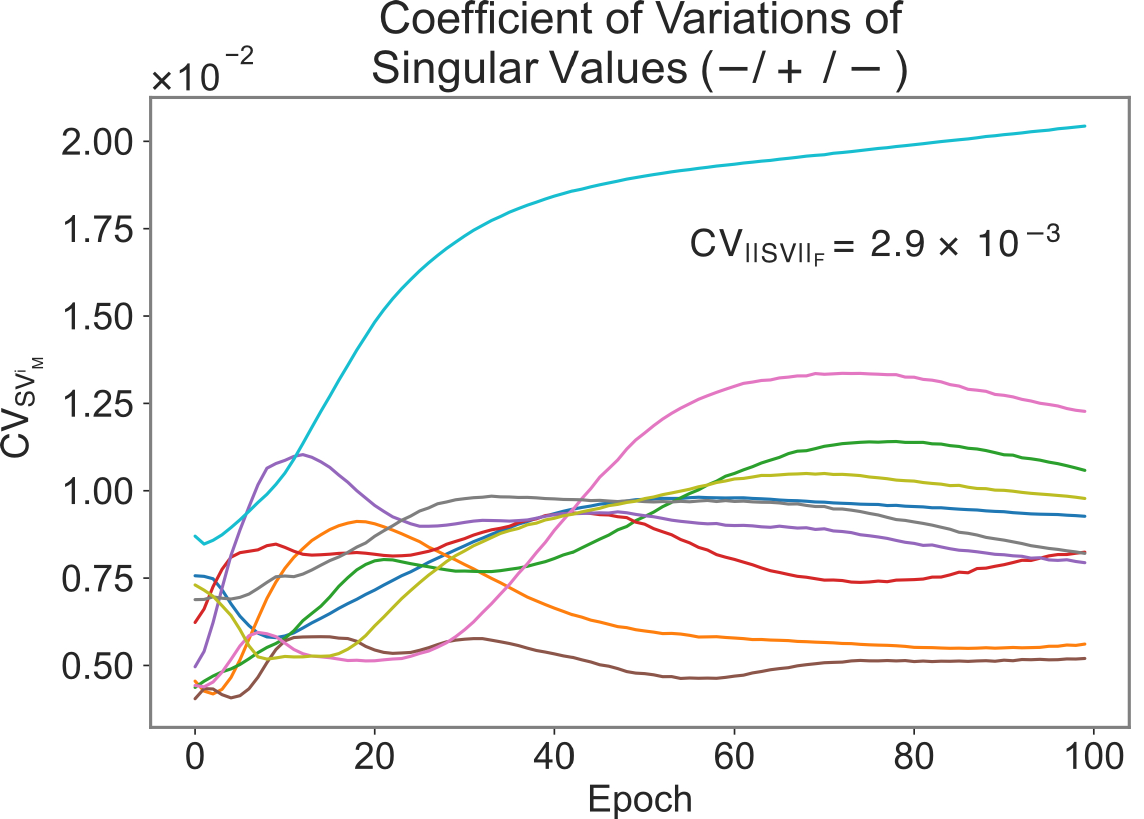}
        \caption{}
        \label{fig:svd_random_100_logit}
    \end{subfigure}
    \hspace{-0.15cm}
    \vspace{0.1cm}
    \begin{subfigure}{0.24\textwidth}
        \includegraphics[width=\textwidth]{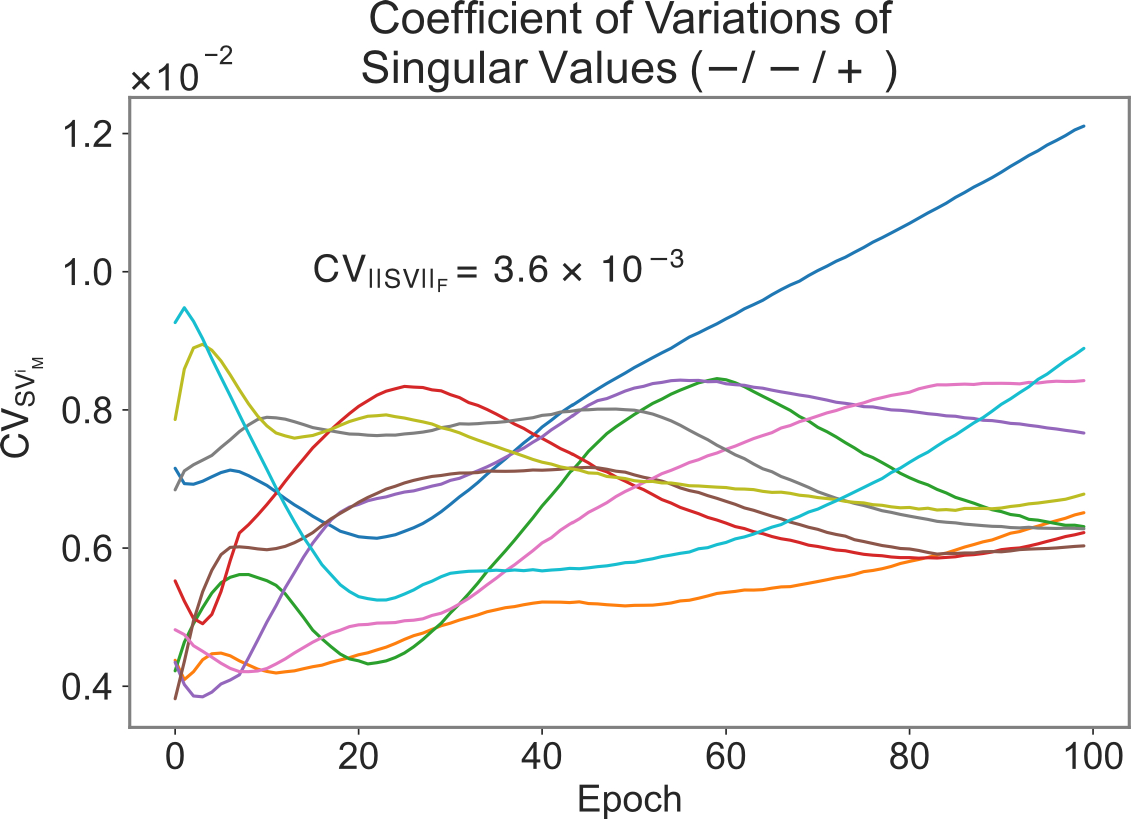}
        \caption{}
        \label{fig:svd_100_label_smoothing}
    \end{subfigure}
    \caption{Network diversity ($\text{CV}_{\text{SV}^i_M}$ and $\text{CV}_{\|\text{SV}\|_F}$) of a LLCM-10 and the MNIST dataset. Panel (a) shows the results of having identically initialized members, whereas panels (b--d) shows the results of random initialization using class weights / logit normalization / label smoothing (amount of smoothing, 0.1). These hyperparameters are either applied (denoted by $+$) or omitted (denoted by $-$).}
    \label{fig:cv_svd_mnist}
\end{figure}

% \FloatBarrier

We investigated identical and non-identical initialization strategies across committee members for a LLCM-10 ($p=0.0)$ network and the MNIST dateset.
In this case, there are a total of 10 singular values for each committee member reflecting the $n=10$ classes of the MNIST dataset.
To gain additional insight during training, for each $\text{SV}^i_m$ we computed the CV across all committee members $m$ ($\text{CV}_{\text{SV}^i_M}$).
We then plotted $\text{CV}_{\text{SV}^i_M}$ to observe the variation of the $\text{SV}^i_M$ during training to confirm if committee members are converging (\autoref{fig:cv_svd_mnist}).
When identically initializing all committee members using the same random seed, we observed no variation of $\text{CV}_{\text{SV}^i_M}$ or $\text{CV}_{\|\text{SV}\|_{F}}$ (\autoref{fig:svd_100}).
However, when committee members are randomly initialized and trained using different hyperparameters such as cross-entropy (\autoref{fig:svd_random_100}) with logit normalization (\autoref{fig:svd_random_100_logit}) or label smoothing (\autoref{fig:svd_100_label_smoothing}), after training we observe ${\text{CV}}_{\|\text{SV}\|_{F}}$ is non-zero and $\text{CV}_{\text{SV}^i_M}$ are non-identical.
In contrast to the $\sigma_{\text{CV}_M}$ analyses of committee members classifiers weights (\autoref{tab:cv_mnist}) where a large range $\sigma_{\text{CV}_M}$ values were obtained, the SVD analyses provided similar results when using different hyperparameters (\textit{e.g.},~\autoref{fig:svd_random_100}--\autoref{fig:svd_100_label_smoothing}: ${\text{CV}}_{\|\text{SV}\|_{F}}~2.7 \times 10^{-3} \text{~--~} 3.6 \times 10^{-3}$).

\subsection{Uncertainty evaluators: A benchmark study}
\label{sect:uncertainity_mnist}

\begin{figure}[!tb]
    \centering
    \begin{subfigure}{0.24\textwidth}
        \includegraphics[width=\textwidth]{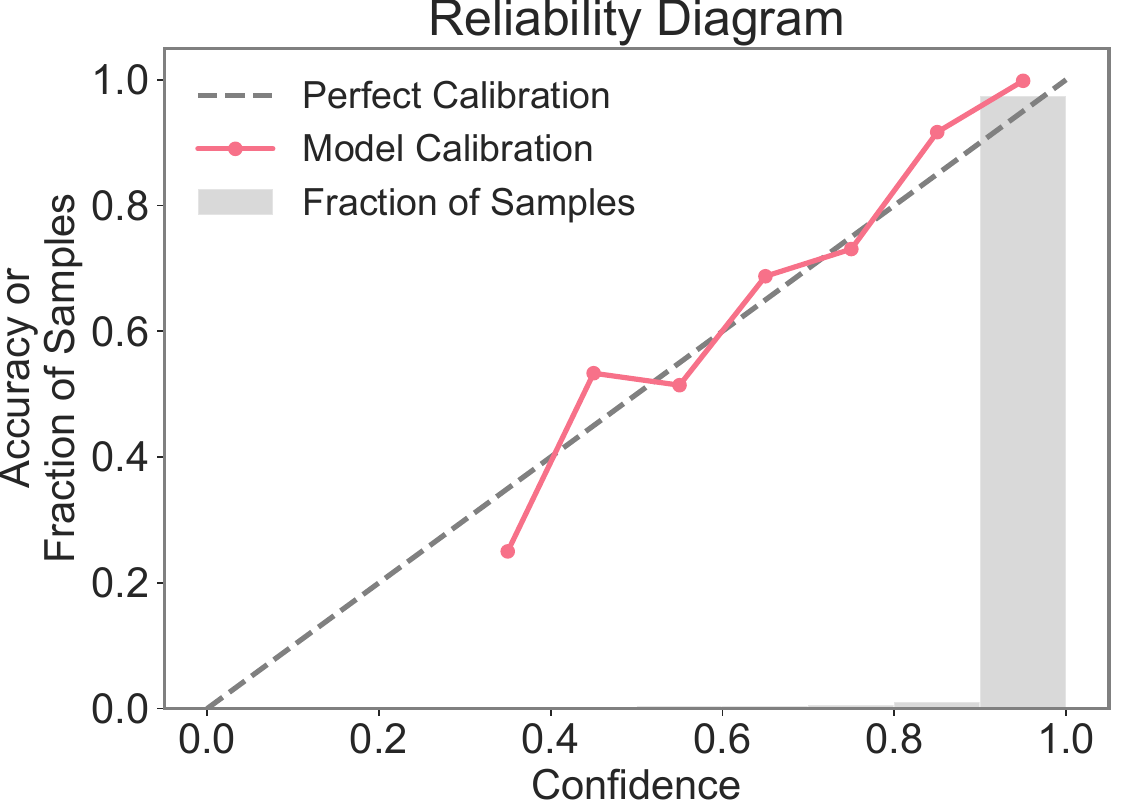}
        \caption{Non-calibrated: $-/-/-$}
        % \label{}
    \end{subfigure}
    \hspace{-0.15cm}
    \vspace{0.25cm}
    \begin{subfigure}{0.24\textwidth}
        \includegraphics[width=\textwidth]{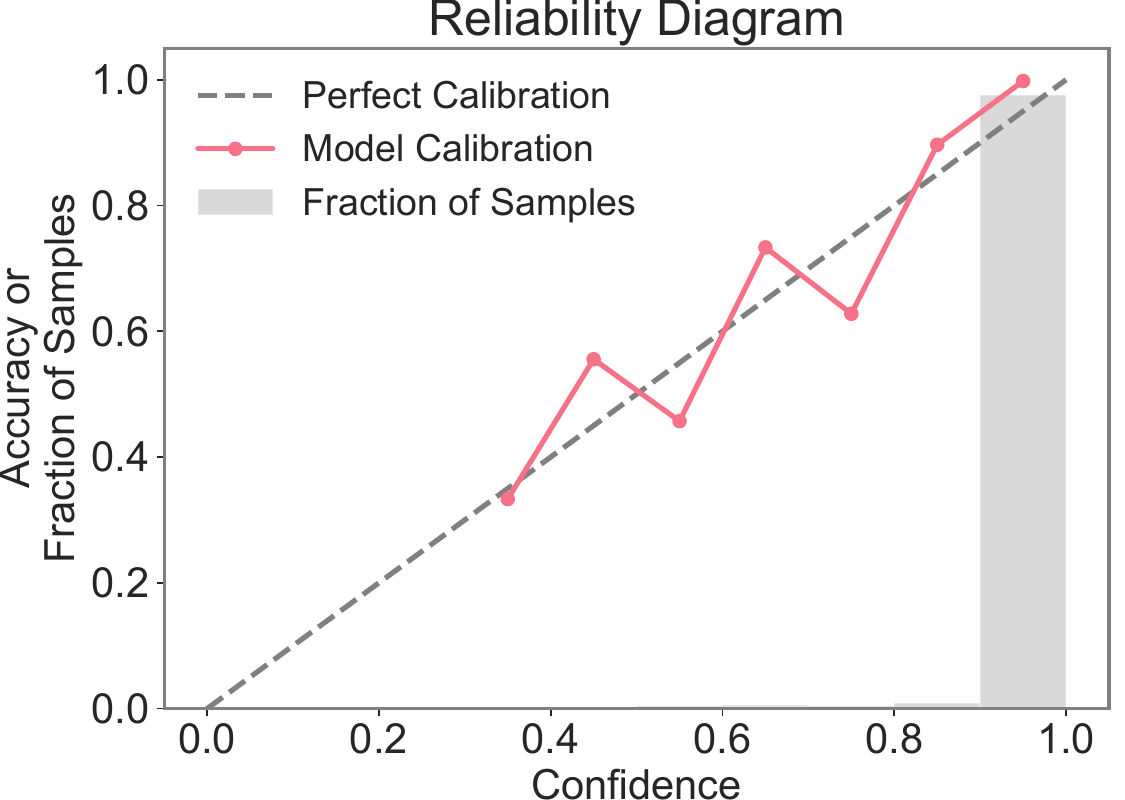}
        \caption{Calibrated: $-/-/-$}
        % \label{}
    \end{subfigure}
    \begin{subfigure}{0.24\textwidth}
        \includegraphics[width=\textwidth]{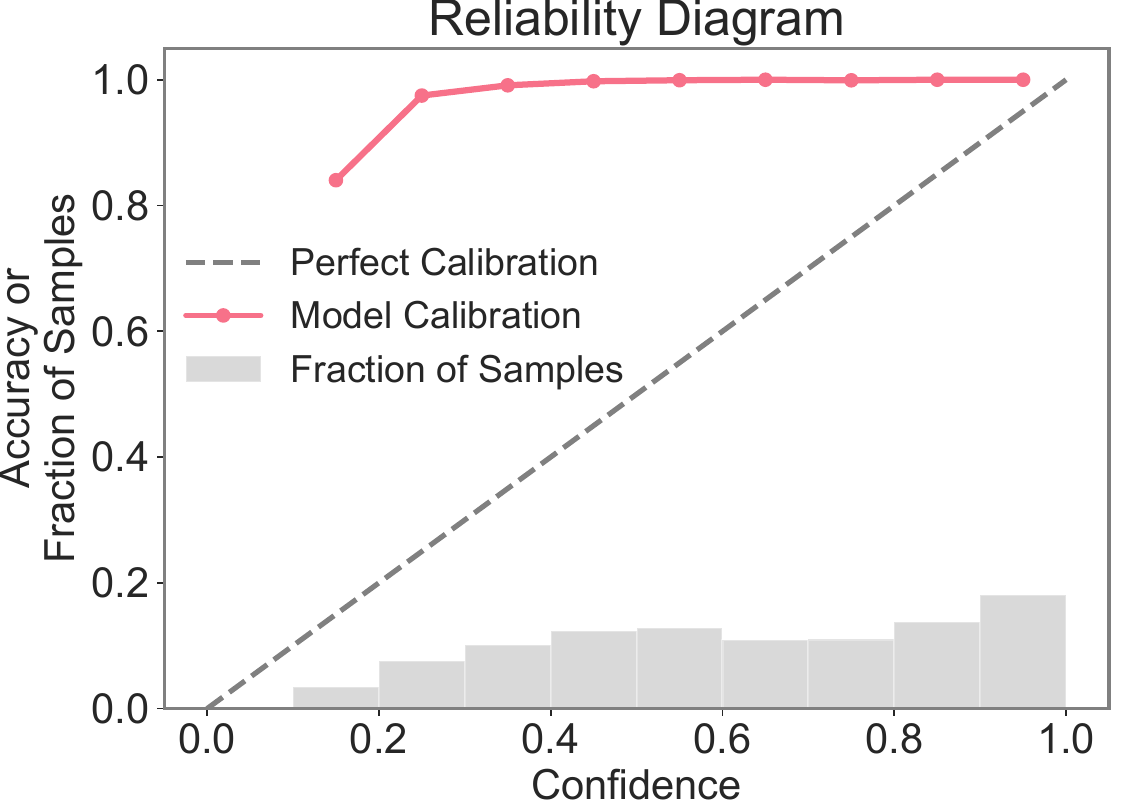}
        \caption{Non-calibrated: $-/+/-$}
        % \label{}
    \end{subfigure}
    \hspace{-0.15cm}
    \vspace{0.25cm}
    \begin{subfigure}{0.24\textwidth}
        \includegraphics[width=\textwidth]{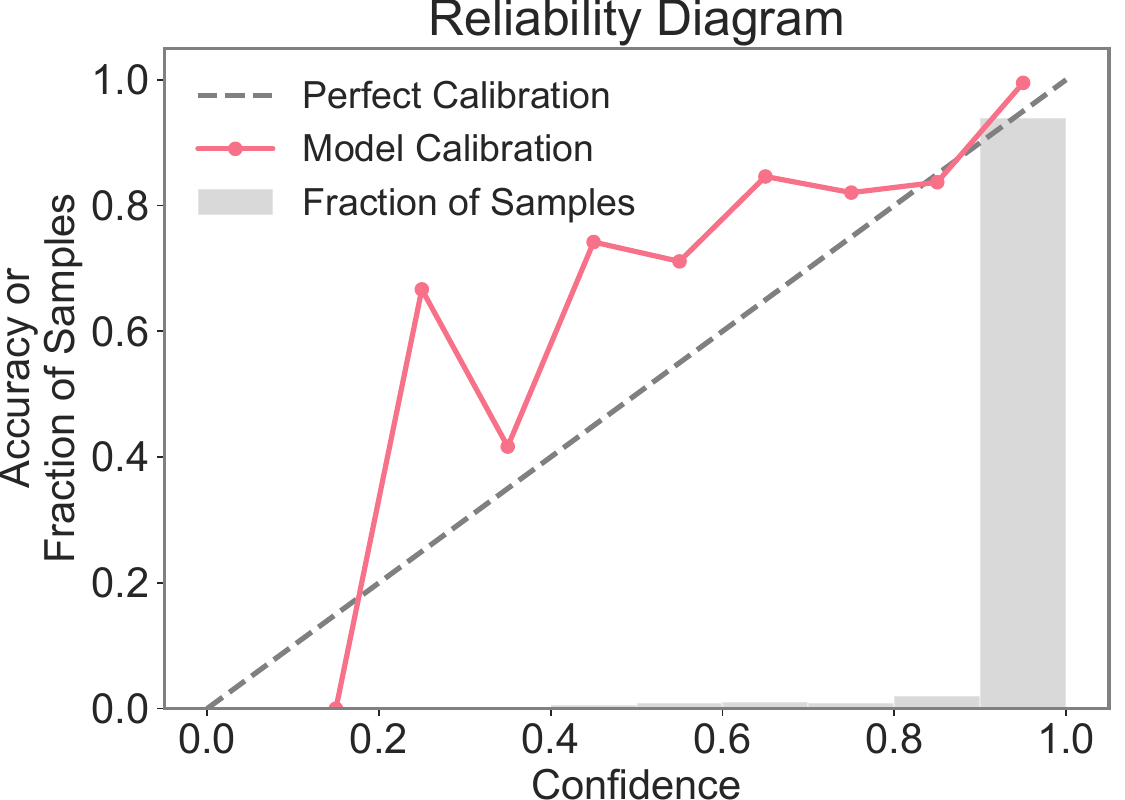}
        \caption{Calibrated: $-/+/-$}
        % \label{}
    \end{subfigure}
    \begin{subfigure}{0.24\textwidth}
        \includegraphics[width=\textwidth]{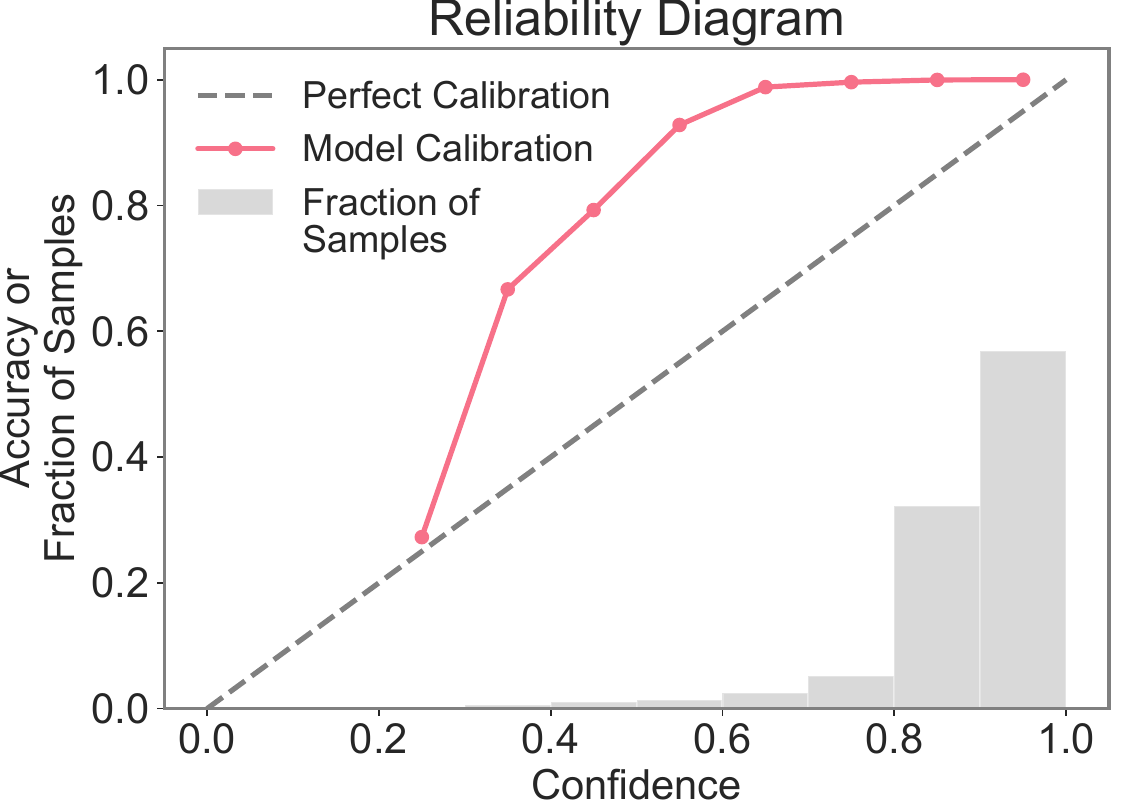}
        \caption{Non-calibrated: $-/-/+$}
        % \label{}
    \end{subfigure}
    \hspace{-0.15cm}
    \vspace{0.25cm}
    \begin{subfigure}{0.24\textwidth}
        \includegraphics[width=\textwidth]{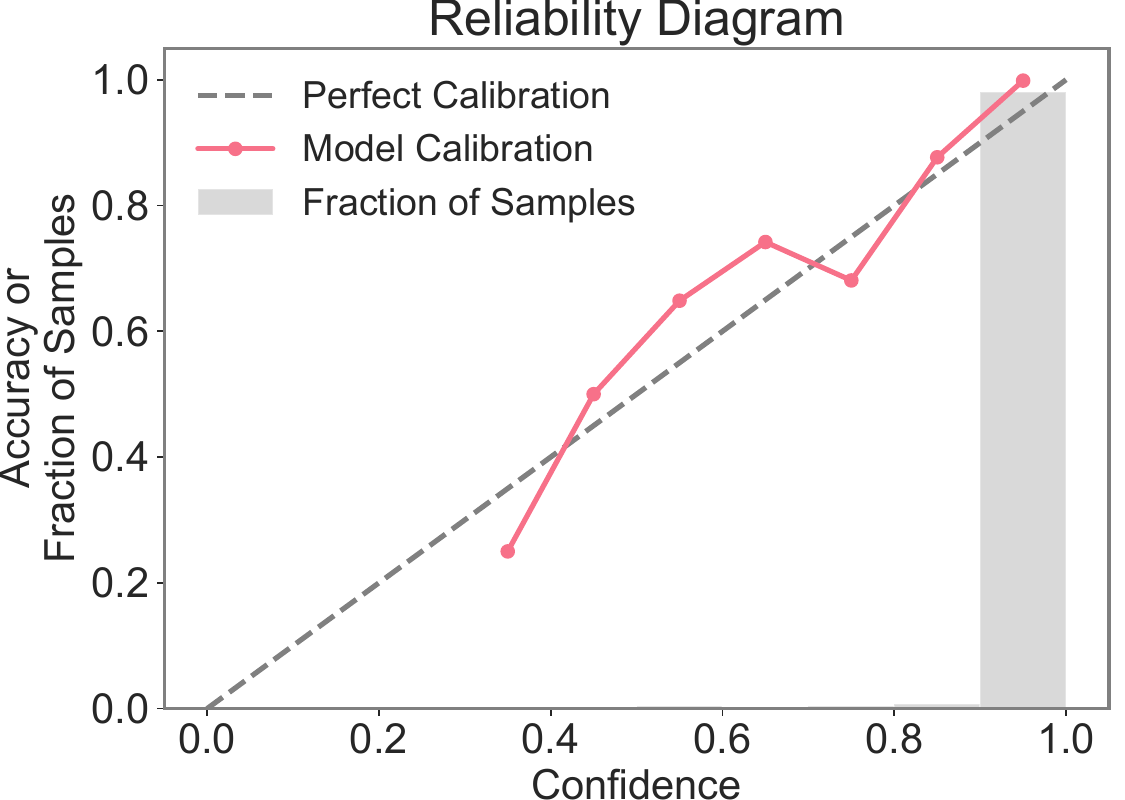}
        \caption{Calibrated: $-/-/+$}
        % \label{}
    \end{subfigure}
    \vspace{-0.75cm}
    \caption{Reliability diagrams for the MNIST dataset using a LLCM-100 network ($p=0.1$). Non-calibrated (left panels) and calibrated (right panels) using class weights / logit normalization / label smoothing (amount of smoothing, 0.1). These hyperparameters are either applied (denoted by $+$) or omitted (denoted by $-$).}
    \label{fig:mnist_reliability}
\end{figure}

For our benchmark study, we used the well-known MNIST dataset and the network configurations as described above (\autoref{sect:methods}).
We first trained a baseline deterministic CNN model with a dropout rate of 0.1 using several combinations of techniques for addressing over-confidence (\autoref{tab:dataset_mnist}, CNN).
When using logit normalization or label smoothing, we observed a noticeable decrease in model calibration as measured by NLL, BS, and ECE.
All of which were resolved after re-calibration by applying a temperature-scaling (see~\autoref{fig:mnist_reliability}).
That is, values tended towards the values obtained without these techniques such as with/without class weighting using the cross-entropy loss function.
This suggests that the networks were already well-calibrated and these steps may not have been required.
However, it would be quite prudent to first measure the amount of network calibration before applying these techniques.
We applied these technique here to gain a overall impression on the effect of results to make generalizations.

\begin{table}[!tb]
\begin{adjustbox}{width=0.48\textwidth}
    \centering
    \begin{threeparttable}
        \caption{Model performances using ensemble and non-ensemble models and the MNIST dataset.}
        \begin{tabular}{ l l l l l l l }
            \toprule
            % headings
              \textbf{Model}\tnote{1,2}
            & \textbf{Accuracy / F1$\uparrow$}
            & \textbf{NLL$\downarrow$}
            & \textbf{BS$\downarrow$}
            & \textbf{ECE$\downarrow$}
            \\
            \midrule
            \multicolumn{5}{l}{\textbf{Non-ensemble:} CNN ($p=0.1$)} \\
            \midrule
            % entry: logs/cnn01/01; logs/cnnp01_calibrated/01
              $-$/$-$/$-$
            & 0.992 / 0.991
            & 0.011 (0.007)
            & 0.003 (0.002)
            & 0.001 (0.003)
            \\
            % entry: logs/cnn01/02; logs/cnnp01_calibrated/02
              $-$/$+$/$-$
            & 0.990 / 0.987
            & 0.457 (0.015)
            & 0.153 (0.004)
            & 0.308 (0.004)
            \\
            % entry: logs/cnn01/03; logs/cnnp01_calibrated/03
              $+$/$+$/$-$
            & 0.990 / 0.987
            & 0.470 (0.016)
            & 0.158 (0.004)
            & 0.316 (0.004)
            \\
            % entry: logs/cnn01/04; logs/cnnp01_calibrated/04
              $+$/$-$/$-$
            & 0.991 / 0.990
            & 0.011 (0.007)
            & 0.003 (0.002)
            & 0.001 (0.003)
            \\
            % entry: logs/cnn01/05; logs/cnnp01_calibrated/05
              $+$/$-$/$+$
            & 0.993 / 0.991
            & 0.141 (0.008)
            & 0.025 (0.002)
            & 0.117 (0.001)
            \\
            % entry: logs/cnn01/06; logs/cnnp01_calibrated/06
              $-$/$-$/$+$
            & 0.991 / 0.991
            & 0.258 (0.009)
            & 0.059 (0.002)
            & 0.208 (0.003)
            \\
            \midrule
            \multicolumn{5}{l}{\textbf{Ensemble: BMA-100}} \\
            \midrule
            % entry: logs/bnn/01
              BNN
            & 0.967 / 0.967
            & 0.164 
            & 0.041 
            & 0.101 
            \\
            \midrule
            \multicolumn{5}{l}{\textbf{Ensemble: MCD-100} (CNN; $p=0.1$)} \\
            \midrule
            % entry: logs/cnnp01/01/dropouts; logs/cnnp01_calibrated/01/dropouts
              $-$/$-$/$-$
            & 0.991 / 0.991
            & 0.016 (0.011)
            & 0.004 (0.003)
            & 0.005 (0.001)
            \\
            % entry: logs/cnnp01/02/dropouts; logs/cnnp01_calibrated/02/dropouts
              $-$/$+$/$-$
            & 0.989 / 0.989
            & 0.419 (0.015)
            & 0.137 (0.004)
            & 0.285 (0.002)
            \\
            % entry: logs/cnnp01/03/dropouts; logs/cnnp01_calibrated/03/dropouts
              $+$/$+$/$-$
            & 0.989 / 0.989
            & 0.431 (0.016)
            & 0.141 (0.004)
            & 0.293 (0.002)
            \\
            % entry: logs/cnnp01/04/dropouts; logs/cnnp01_calibrated/04/dropouts
              $+$/$-$/$-$
            & 0.991 / 0.991
            & 0.015 (0.011)
            & 0.004 (0.003)
            & 0.004 (0.002)
            \\
            % entry: logs/cnnp01/05/dropouts; logs/cnnp01_calibrated/05/dropouts
              $+$/$-$/$+$
            & 0.993 / 0.993
            & 0.140 (0.010)
            & 0.025 (0.002)
            & 0.115 (0.001)
            \\
            % entry: logs/cnnp01/06/dropouts; logs/cnnp01_calibrated/06/dropouts
              $-$/$-$/$+$
            & 0.992 / 0.991
            & 0.365 (0.055)
            & 0.099 (0.010)
            & 0.287 (0.039)
            \\
            \midrule
            \multicolumn{5}{l}{\textbf{Ensemble: LLCM-100} (LLCM; $M=100;~p=0.1$)} \\
            \midrule
            % entry: logs/multicnn100p01/01; logs/multicnn100p01_calibrated/01
              $-$/$-$/$-$
            & 0.992 / 0.992
            & 0.011 (0.009)
            & 0.003 (0.002)
            & 0.002 (0.002)
            \\
            % entry: logs/multicnn100p01/02; logs/multicnn100p01_calibrated/02
              $-$/$+$/$-$
            & 0.991 / 0.991
            & 0.558 (0.015)
            & 0.196 (0.004)
            & 0.362 (0.005)
            \\
            % entry: logs/multicnn100p01/03; logs/multicnn100p01_calibrated/03
              $+$/$+$/$-$
            &  0.991 / 0.991
            &  0.539 (0.015)
            &  0.188 (0.004)
            &  0.352 (0.005)
            \\
            % entry: logs/multicnn100p01/04; logs/multicnn100p01_calibrated/04
              $+$/$-$/$-$
            & 0.993 / 0.993
            & 0.012 (0.009)
            & 0.003 (0.002)
            & 0.003 (0.002)
            \\
            % entry: logs/multicnn100p01/05; logs/multicnn100p01_calibrated/05
              $+$/$-$/$+$
            & 0.993 / 0.993
            & 0.141 (0.008)
            & 0.024 (0.002)
            & 0.117 (0.001)
            \\
            % entry: logs/multicnn100p01/06; logs/multicnn100p01_calibrated/06
              $-$/$-$/$+$
            & 0.994 / 0.993
            & 0.139 (0.008)
            & 0.024 (0.002)
            & 0.116 (0.001)
            \\
            \midrule
            \bottomrule
        \end{tabular}
        % add notes using \tnote{#}
        $^1$ BMA: 100 models; MCD: 100 inference samplings; and LLCM: 100 committee members. 
        $^2$ Models defined based on training hyperparameters using class weights / logit normalization / label smoothing (amount of smoothing, 0.1). These hyperparameters are either applied (denoted by $+$) or omitted (denoted by $-$). Values in parentheses are for calibrated models using calculated temperature(s) as previously described. F1-scores reported as macro-averaged.
        \label{tab:dataset_mnist}
    \end{threeparttable}
\end{adjustbox}
\end{table}

% \FloatBarrier

We now compare the different uncertainty evaluators (BMA, MCD, LLCM) for the MNIST dataset.
We converted the CNN network to a BNN ($p=0.0$) using the tools available from the Pyro framework.
This BNN was then trained and BMA was performed by sampling 100 sets of weights from the learned distributions and evaluating the resulting ensemble (\autoref{tab:dataset_mnist}, BMA-100).
While the overall performance was slightly less than the deterministic model (\textit{ca.} 97\% \textit{vs.} 99\%) this was not necessarily unexpected given that a different model, loss objective (\textit{i.e.}, ELBO), and a dropout layer were used for the deterministic model.
Monte Carlo dropout inference sampling was then performed using a trained CNN ($p=0.1$) with an inference dropout rate of $0.1$, sampling 100 sets of weights, and evaluating the resulting ensemble (\autoref{tab:dataset_mnist}, MCD-100).
As with the baseline CNN, we explored different techniques for addressing over-confidence, where the results were comparable to the CNN.
Subsequently, we trained our LLCM model using 100 committee members with a dropout rate of $0.1$ (\autoref{tab:dataset_mnist}, LLCM-100).
In this case, sampling 100 sets of weights was not required as a single forward-pass returns per-sample uncertainties from the 100 committee members.
For the BMA, MCD and LLCM networks, all corresponding models were comparable to the baseline CNN.
However, the advantage of the LLCM compared to BMA or MCD is the reduced compute requirements without any performance loss.

For example, in this case, the CNN network used for MCD consists of \textit{ca.} 400K network parameters which is sampled $100 \times$ (\textit{i.e.}, 40M network parameters) to obtain per-sample uncertainties. 
Whereas, the LLCM network consists of \textit{ca.} 2M network parameters which is evaluated only once.
This represents a $20 \times$ reduction (95\%) of network parameters to obtain per-sample uncertainties.

\begin{figure}[!htb]
    \centering
    \begin{subfigure}{0.24\textwidth}
        \includegraphics[width=\textwidth]{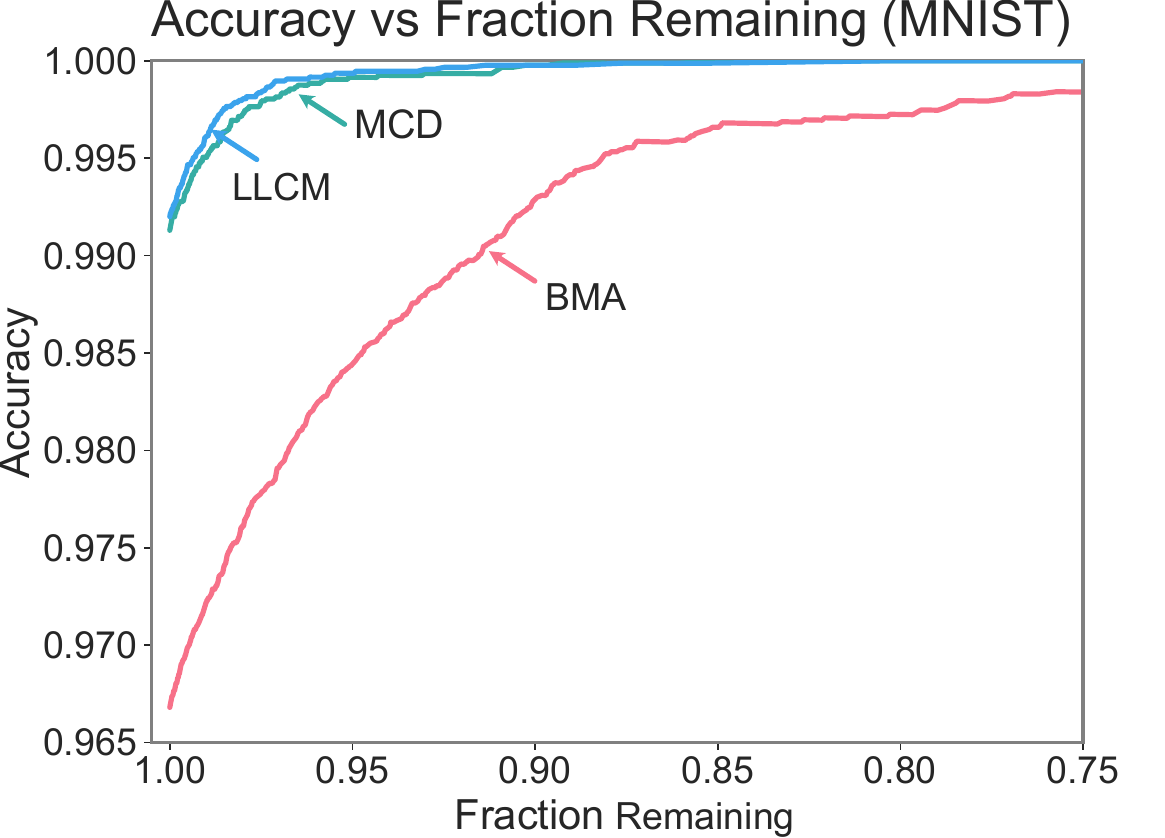}
        \caption{}
        \label{fig:mnist_p_scores_accuracy}
    \end{subfigure}
    \hspace{-0.3cm}
    \begin{subfigure}{0.24\textwidth}
        \includegraphics[width=\textwidth]{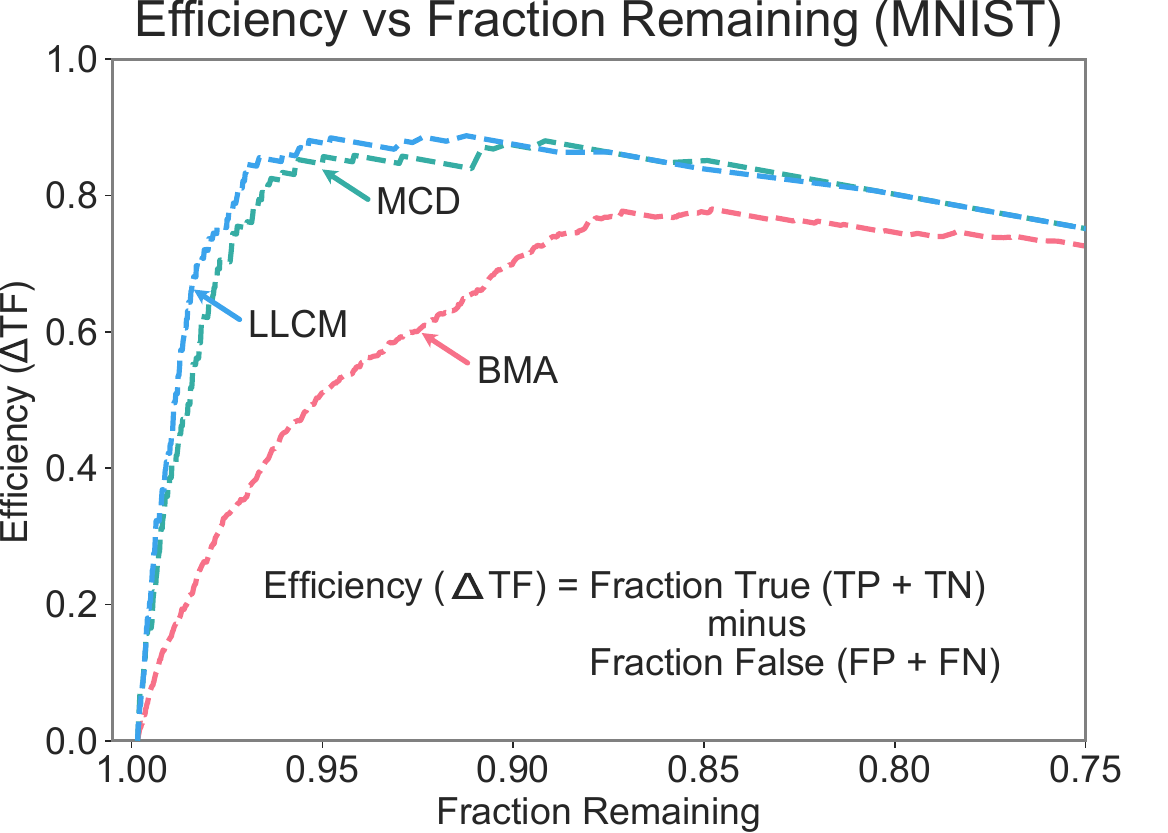}
        \caption{}
    \label{fig:mnist_p_scores_efficiency}
    \end{subfigure}
    \vspace{-0.1cm}
    \caption{Uncertain metric plots for the BMA-100, MCD-100 ($-/-/-$), and LLCM-100 ($-/-/-$) networks and the MNIST dataset using class weights / logit normalization / label smoothing (amount of smoothing, 0.1). These hyperparameters are either applied (denoted by $+$) or omitted (denoted by $-$). Each trace is created by applying increasing thresholds (step size 0.001) of confidence values and re-calculating metrics.}
    \label{fig:mnist_p_scores}
\end{figure}

To compare the uncertainty estimations of BMA, MCD and LLCM networks, we plotted uncertain metric plots (\autoref{fig:mnist_p_scores}).
For illustration, we show the results using cross-entropy loss without logit normalization and label smoothing (\textit{i.e.}, $-/-/-$).
Both the MCD and LLCM provided similar results followed by BMA.
\autoref{fig:mnist_p_scores_efficiency} shows there is a preferential removal of incorrect model predictions (FP + FN) as opposed to correct model predictions (TP + TN) up to an optimal fraction.
Recall that the fraction of remaining samples is calculated based on different threshold of confidence values.
After this fraction, the differences of the removal rates are decreasing and do not offer additional benefits.
In fact, performance is artificially enhances by the non-preferential removal of incorrect predictions.

% \FloatBarrier

% \FloatBarrier

In summary, the benchmark study demonstrates that LLCMs can further simplify last-head strategies (see~\autoref{sect:related}) by only using a single last-layer for uncertainty estimations.
Both MCD and LLCM approaches resulted with similar performances (with slightly better performance than BMA); however, LLCMs dramatically reduces the computational requirements.
Applying over-confident techniques such as logit normalization or label smoothing appeared to degrade model calibration, which can be corrected using temperature-scaling.
Overall, this offers an efficient approach for Bayesian approximations and a strategy to identify uncertain model predictions for human-in-the-loop interventions.

\subsection{Uncertainty evaluators: Benthic imagery}
\label{sect:uncertainity_benthic}

\paragraph{German Bank 2010 dataset}
The German Bank dataset consists of 5 classes of silt/mud, silt with bedforms, reef, glacial till, and sand with bedforms, and represents a difficult dataset for learning a model.
The results of the dataset is summarized in \autoref{tab:dataset_gb}.
We initially trained a non-ensemble CNN (ResNet-CNN) with a dropout rate of $0.01$ from a pre-trained model, using class weighting and/or different techniques for addressing over-confidence (\autoref{tab:dataset_gb}, Non-ensemble).
We obtained modest performance for accuracy and calibration metrics as observed from accuracies, NLL, BS, and ECE.
In fact, model calibration decreased when using techniques for addressing over-confidence (\textit{e.g.}, logit normalization and label smoothing).
A variational Bayesian last-layer network based on the pre-trained ResNet-CNN model was subsequently created and trained.
After performing BMA with 100 samplings of weights, the performance was slightly less than the non-ensemble model (\autoref{tab:dataset_gb}, BMA-100 \textit{vs.} Non-ensemble).

\begin{table}[!tb]
    \centering
    \begin{adjustbox}{width=0.48\textwidth}
    \begin{threeparttable}
        \caption{Model performances using ensemble and non-ensemble models and the German Bank 2010 dataset.}
        \begin{tabular}{ l l l l l l }
            \toprule
            % headings
              \textbf{Model}\tnote{1,2}
            & \textbf{Accuracy / F1$\uparrow$}
            & \textbf{NLL$\downarrow$}
            & \textbf{BS$\downarrow$}
            & \textbf{ECE$\downarrow$}
            \\
            \midrule
            \multicolumn{5}{l}{\textbf{Non-ensemble:} ResNet-CNN ($p=0.01$)} \\
            \midrule
            % entry: logs/german_bank_2010/resnet50p01bt/01; logs/german_bank_2010/resnet50p01bt_calibrated/01
              $-$/$-$/$-$
            & 0.790 / 0.738
            & 0.202 (0.200)
            & 0.058 (0.057)
            & 0.062 (0.060)
            \\
            % entry: logs/german_bank_2010/resnet50p01bt/02; logs/german_bank_2010/resnet50p01bt_calibrated/02
              $-$/$+$/$-$
            & 0.770 / 0.711
            & 0.871 (0.201)
            & 0.334 (0.060)
            & 0.328 (0.082)
            \\
            % entry: logs/german_bank_2010/resnet50p01bt/03; logs/german_bank_2010/resnet50p01bt_calibrated/03
              $+$/$+$/$-$
            & 0.758 / 0.700
            & 0.866 (0.198)
            & 0.331 (0.059)
            & 0.314 (0.101)
            \\
            % entry: logs/german_bank_2010/resnet50p01bt/04; logs/german_bank_2010/resnet50p01bt_calibrated/04
              $+$/$-$/$-$
            & 0.794 / 0.744
            & 0.202 (0.188)
            & 0.057 (0.053)
            & 0.069 (0.075)
            \\
            % entry: logs/german_bank_2010/resnet50p01bt/05; logs/german_bank_2010/resnet50p01bt_calibrated/05
              $+$/$-$/$+$
            & 0.786 / 0.734
            & 0.338 (0.189)
            & 0.101 (0.052)
            & 0.058 (0.077)
            \\
            % entry: logs/german_bank_2010/resnet50p01bt/06; logs/german_bank_2010/resnet50p01bt_calibrated/06
              $-$/$-$/$+$
            & 0.774 / 0.717
            & 0.298 (0.196)
            & 0.088 (0.055)
            & 0.023 (0.070)
            \\
            \midrule
            \multicolumn{5}{l}{\textbf{Ensemble: BMA-100} (ResNet-BNN)} \\
            \midrule
            % entry: logs/german_bank_2010/bnnresnet50bt/01
              BNN
            & 0.734 / 0.691
            & 0.527 
            & 0.179 
            & 0.117 
            \\
            \midrule
            \multicolumn{5}{l}{\textbf{Ensemble: MCD-100} (ResNet-CNN; $p=0.01$)} \\
            \midrule
            % entry: logs/german_bank_2010/resnet50p01bt/01/dropouts; logs/german_bank_2010/resnet50p01bt_calibrated/01/dropouts
              $-$/$-$/$-$
            & 0.794 / 0.745
            & 0.225 (0.222)
            & 0.065 (0.064)
            & 0.059 (0.059)
            \\
            % entry: logs/german_bank_2010/resnet50p01bt/02/dropouts; logs/german_bank_2010/resnet50p01bt_calibrated/02/dropouts
              $-$/$+$/$-$
            & 0.772 / 0.717
            & 0.917 (0.232)
            & 0.355 (0.069)
            & 0.351 (0.075)
            \\
            % entry: logs/german_bank_2010/resnet50p01bt/03/dropouts; logs/german_bank_2010/resnet50p01bt_calibrated/03/dropouts
              $+$/$+$/$-$
            & 0.762 / 0.708
            & 0.910 (0.227)
            & 0.352 (0.068)
            & 0.339 (0.070)
            \\
            % entry: logs/german_bank_2010/resnet50p01bt/04/dropouts; logs/german_bank_2010/resnet50p01bt_calibrated/04/dropouts
              $+$/$-$/$-$
            & 0.788 / 0.738
            & 0.226 (0.210)
            & 0.065 (0.059)
            & 0.037 (0.057)
            \\
            % entry: logs/german_bank_2010/resnet50p01bt/05/dropouts; logs/german_bank_2010/resnet50p01bt_calibrated/05/dropouts
              $+$/$-$/$+$
            & 0.784 / 0.733
            & 0.370 (0.213)
            & 0.112 (0.060)
            & 0.068 (0.066)
            \\
            % entry: logs/german_bank_2010/resnet50p01bt/06/dropouts; logs/german_bank_2010/resnet50p01bt_calibrated/06/dropouts
              $-$/$-$/$+$
            & 0.788 / 0.738
            & 0.330 (0.223)
            & 0.099 (0.063)
            & 0.044 (0.053)
            \\
            \midrule
            \multicolumn{5}{l}{\textbf{Ensemble: LLCM-100} (ResNet-LLCM; $M=100;~p=0.01$)} \\
            \midrule
            % entry: logs/german_bank_2010/multiresnet50p01bt/01; logs/german_bank_2010/multiresnet50p01bt_calibrated/01
              $-$/$-$/$-$
            & 0.786 / 0.719
            & 0.151 (0.219)
            & 0.040 (0.059)
            & 0.099 (0.058)
            \\
            % entry: logs/german_bank_2010/multiresnet50p01bt/02; logs/german_bank_2010/multiresnet50p01bt_calibrated/02
              $-$/$+$/$-$
            & 0.774 / 0.691
            & 0.636 (0.255)
            & 0.227 (0.081)
            & 0.207 (0.074)
            \\
            % entry: logs/german_bank_2010/multiresnet50p01bt/03; logs/german_bank_2010/multiresnet50p01bt_calibrated/03
              $+$/$+$/$-$
            & 0.758 / 0.687
            & 0.651 (0.265)
            & 0.233 (0.085)
            & 0.200 (0.052)
            \\
            % entry: logs/german_bank_2010/multiresnet50p01bt/04; logs/german_bank_2010/multiresnet50p01bt_calibrated/04
              $+$/$-$/$-$
            & 0.788 / 0.722
            & 0.145 (0.199)
            & 0.037 (0.052)
            & 0.099 (0.060)
            \\
            % entry: logs/german_bank_2010/multiresnet50p01bt/05; logs/german_bank_2010/multiresnet50p01bt_calibrated/05
              $+$/$-$/$+$
            & 0.788 / 0.728
            & 0.301 (0.214)
            & 0.083 (0.055)
            & 0.047 (0.046)
            \\
            % entry: logs/german_bank_2010/multiresnet50p01bt/06; logs/german_bank_2010/multiresnet50p01bt_calibrated/06
              $-$/$-$/$+$
            & 0.782 / 0.713
            & 0.252 (0.208)
            & 0.067 (0.054)
            & 0.054 (0.060)
            \\
            \midrule
            \bottomrule
        \end{tabular}
        % add notes using \tnote{#}
        $^1$ BMA: 100 models; MCD: 100 inference samplings; and LLCM: 100 committee members. 
        $^2$ Models defined based on training hyperparameters using class weights / logit normalization / label smoothing (amount of smoothing, 0.1). These hyperparameters are either applied (denoted by $+$) or omitted (denoted by $-$). Values in parentheses are for calibrated models using calculated temperature(s) as previously described. F1-scores reported as macro-averaged.
        \label{tab:dataset_gb}
    \end{threeparttable}
    \end{adjustbox}
\end{table}

% \FloatBarrier

\noindent
In this case, the calibration metrics suggests that the model is poorly calibrated.
We then performed Monte Carlo dropout inference sampling on the trained ResNet-CNN models (\autoref{tab:dataset_gb}, Non-ensemble) by sampling 100 sets of weights during inference, which resulted with comparable metrics to the non-ensemble models (\autoref{tab:dataset_gb}, MCD-100 \textit{vs.}~Non-ensemble).
A similar trend for model calibrations were observed for the non-ensemble models.
Lastly, we applied our LLCM by adding an LLCM-100 module to the ResNet-CNN network (\autoref{tab:dataset_gb}, LLCM-100).
After training and a single forward-pass of inputs during inference, we obtained metrics that were comparable to both MCD-100, BMA-100, and non-ensembles models (with the exception of uncertainty metrics for BMA-100).

\begin{figure}[!tb]
    \centering
    \begin{subfigure}{0.24\textwidth}
        \includegraphics[width=\textwidth]{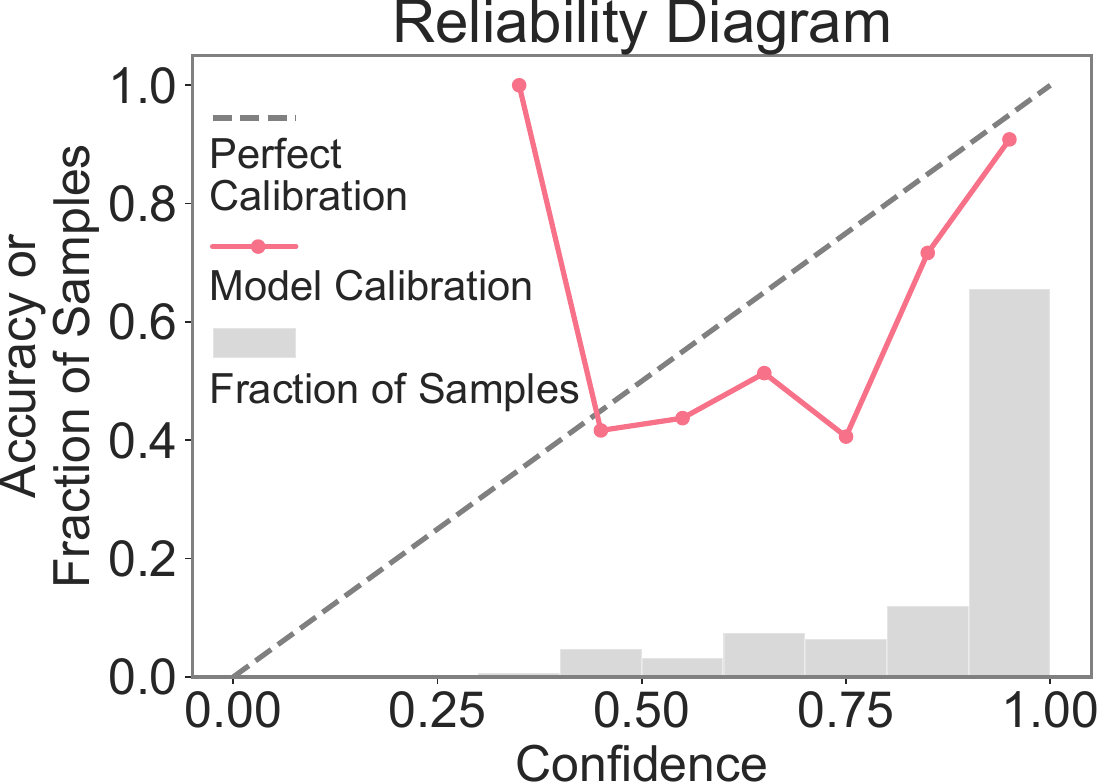}
        \caption{Non-calibrated: $-/-/-$}
        % \label{}
    \end{subfigure}
    \hspace{-0.15cm}
    \vspace{0.25cm}
    \begin{subfigure}{0.24\textwidth}
        \includegraphics[width=\textwidth]{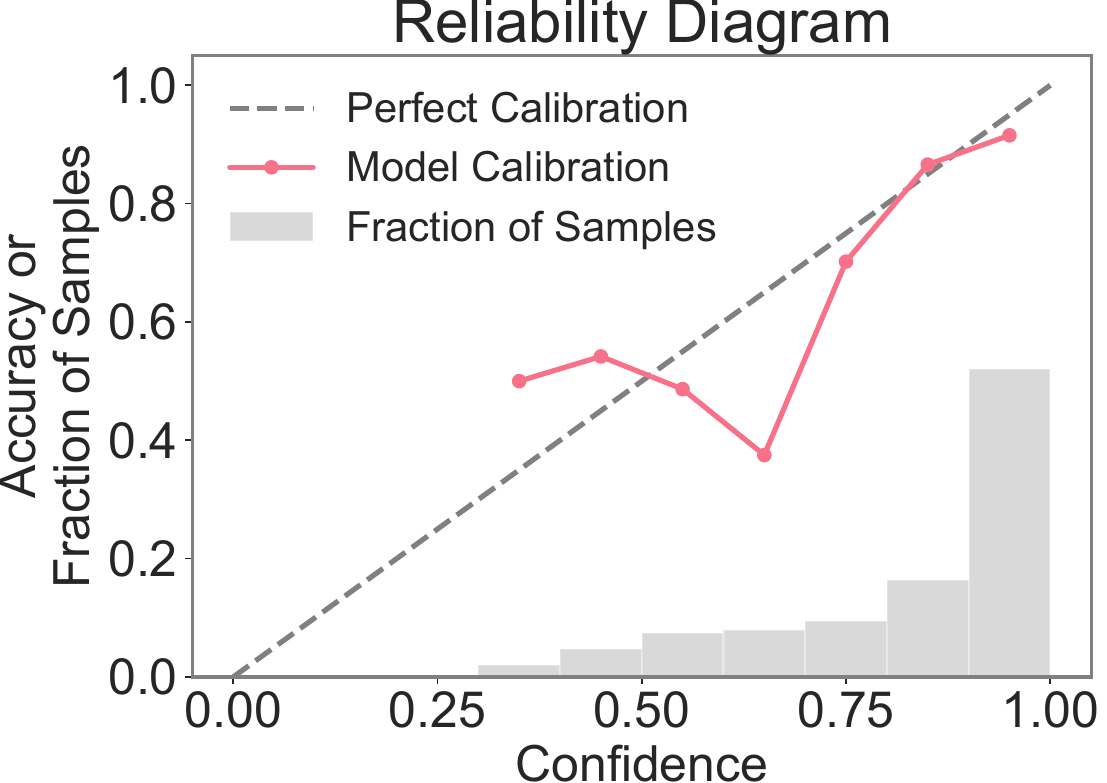}
        \caption{Calibrated: $-/-/-$}
        % \label{}
    \end{subfigure}
    \begin{subfigure}{0.24\textwidth}
        \includegraphics[width=\textwidth]{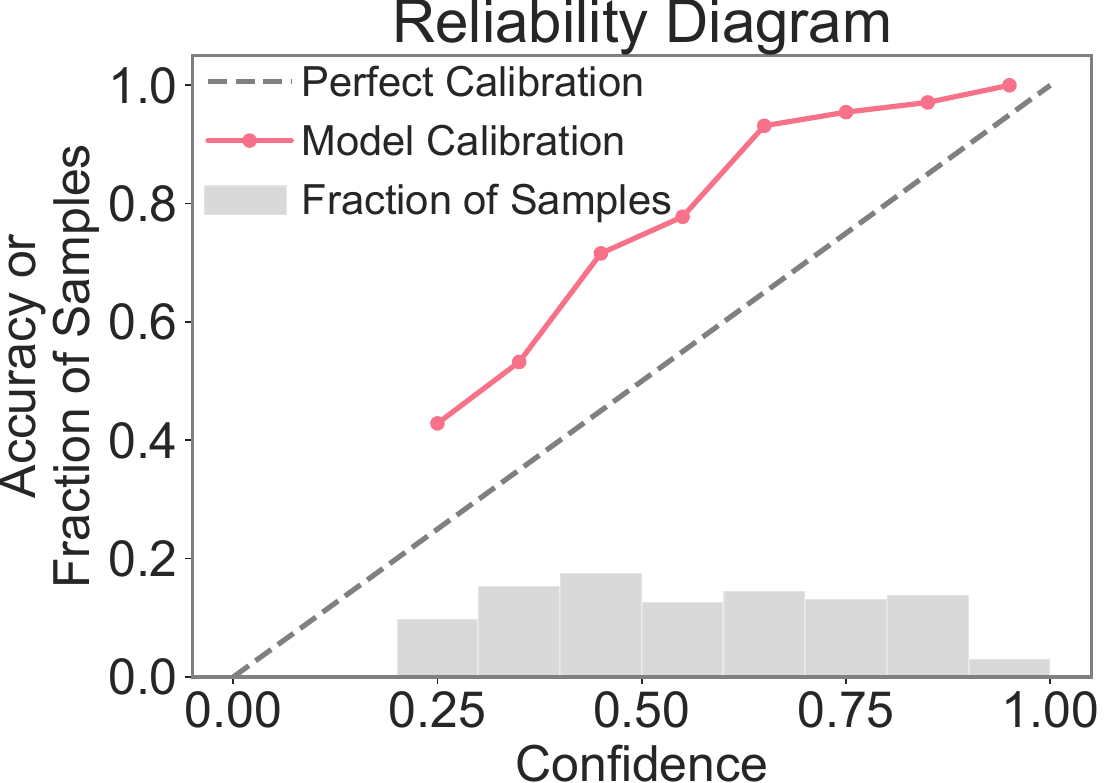}
        \caption{Non-calibrated: $-/+/-$}
        % \label{}
    \end{subfigure}
    \hspace{-0.15cm}
    \vspace{0.25cm}
    \begin{subfigure}{0.24\textwidth}
        \includegraphics[width=\textwidth]{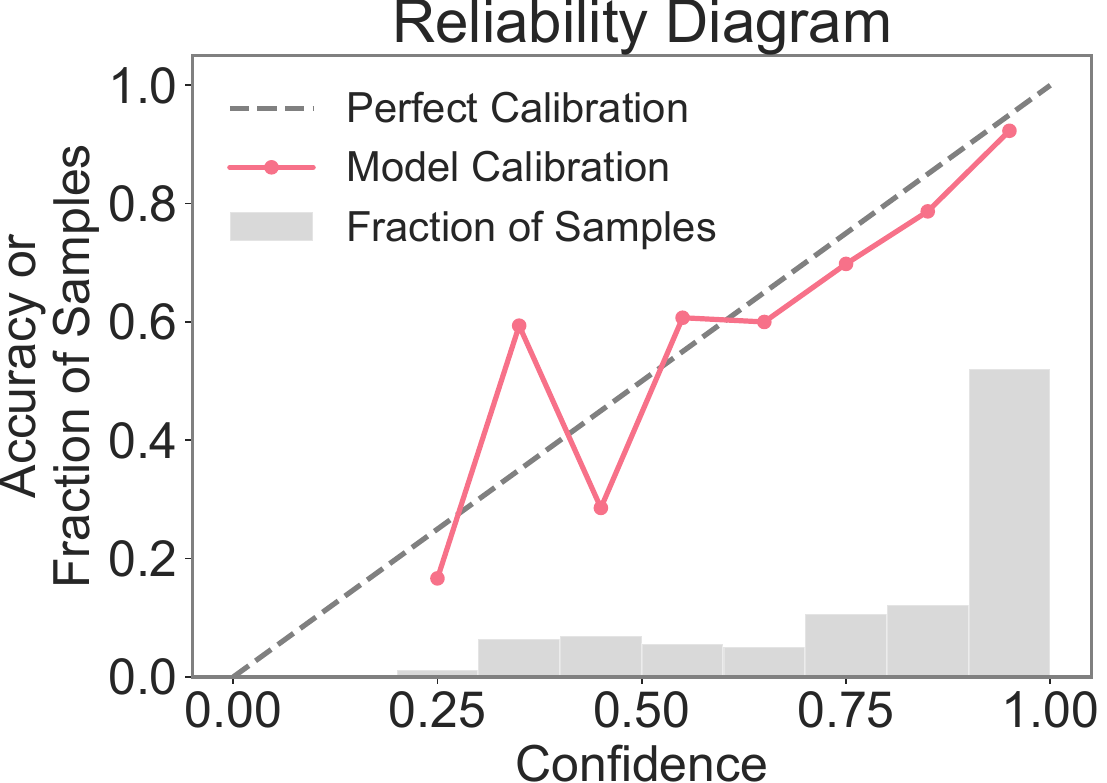}
        \caption{Calibrated: $-/+/-$}
        \label{fig:gb_reliability_logit_calibrated}
    \end{subfigure}
    \begin{subfigure}{0.24\textwidth}
        \includegraphics[width=\textwidth]{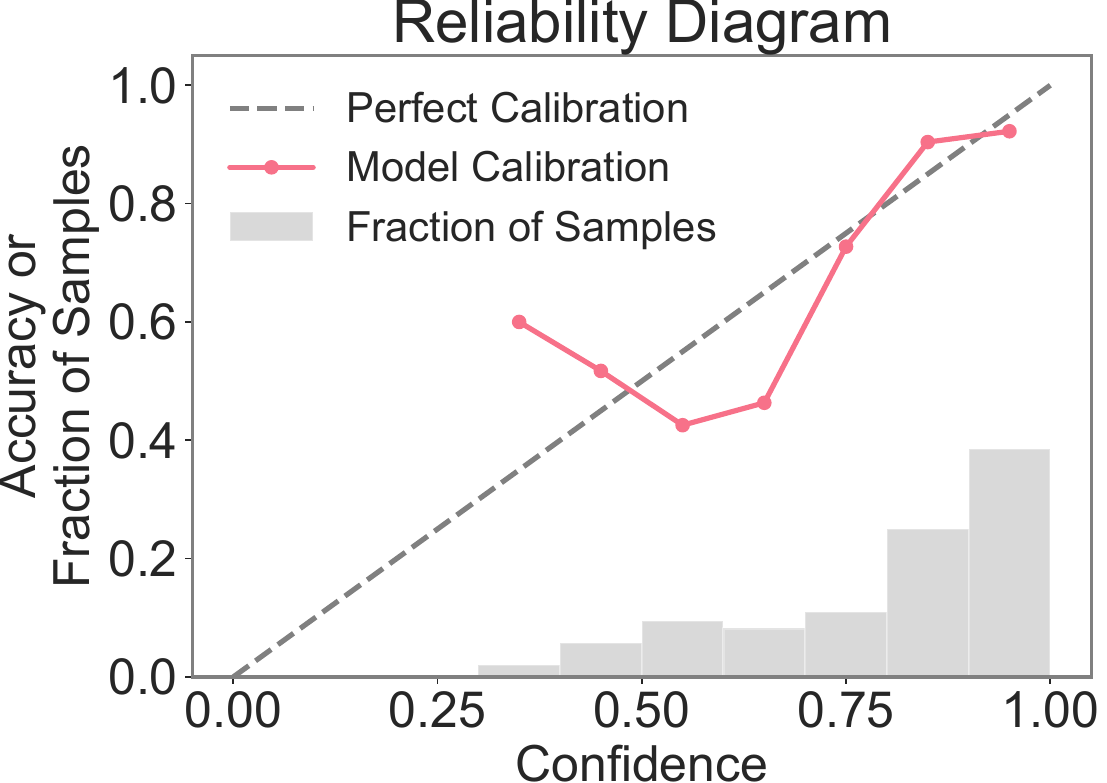}
        \caption{Non-calibrated: $-/-/+$}
        % \label{}
    \end{subfigure}
    \hspace{-0.15cm}
    \vspace{0.25cm}
    \begin{subfigure}{0.24\textwidth}
        \includegraphics[width=\textwidth]{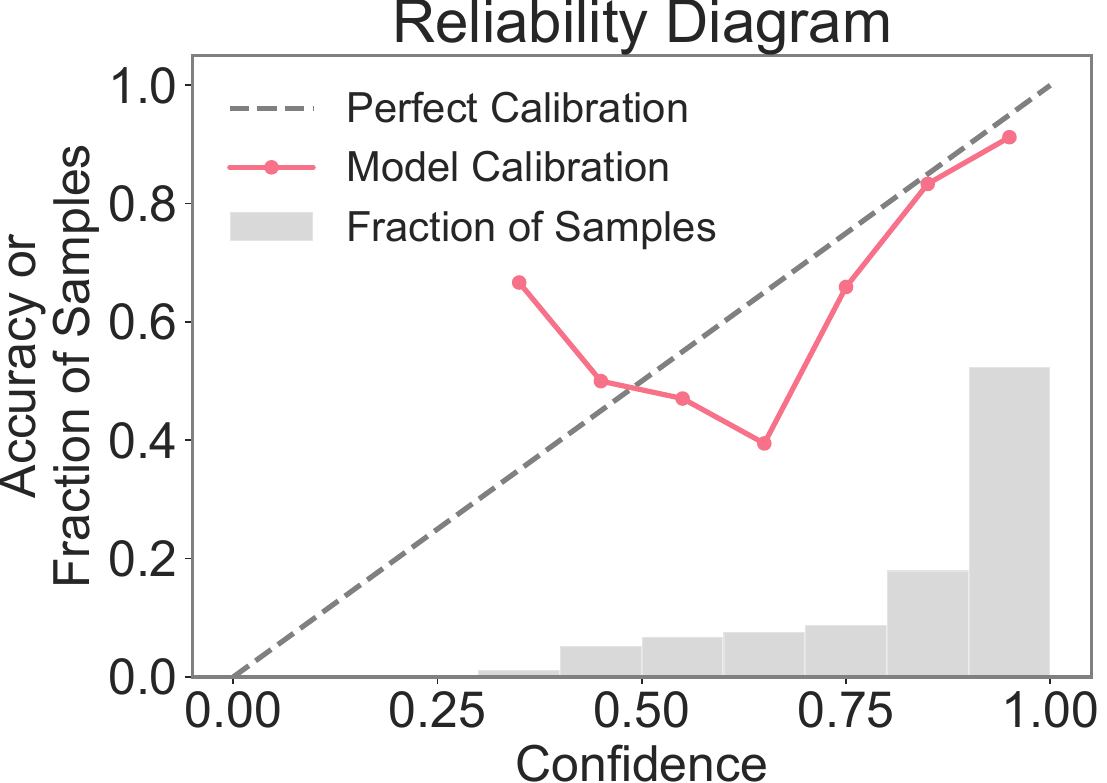}
        \caption{Calibrated: $-/-/+$}
        % \label{}
    \end{subfigure}
    \vspace{-0.75cm}
    \caption{Reliability diagrams for the German Bank 2010 dataset using a LLCM-100 network ($p=0.01$). Non-calibrated (left panels) and calibrated (right panels) using class weights / logit normalization / label smoothing (amount of smoothing, 0.1). These hyperparameters are either applied (denoted by $+$) or omitted (denoted by $-$).}
    \label{fig:gb_reliability}
\end{figure}

% \FloatBarrier

When using techniques for addressing over-confidence~\citep{Szegedy::2015a,Wei::2022a} with the non-ensemble networks and the MCD-100 and LLCM-100 ensemble networks (\autoref{tab:dataset_gb}), we observed an appreciable decrease in model calibration. 
In \autoref{fig:gb_reliability} we plotted reliability diagrams for the LLCM-100 non-calibrated and calibrated models using cross-entropy loss with ($-/+/-$) or without ($-/-/-$) logit normalization or label smoothing ($-/-/+$).
The largest effect observed involved the use of logit normalization, where the model was under-confident with a near uniform distribution of confidence values.
However, this was mostly restored after applying a re-calibration (\autoref{fig:gb_reliability_logit_calibrated}).
This was also observed in our benchmark study for both logit normalization and label smoothing (\autoref{fig:mnist_reliability}).

\begin{figure}[!tb]
    \centering
    \begin{subfigure}{0.24\textwidth}
        \includegraphics[width=\textwidth]{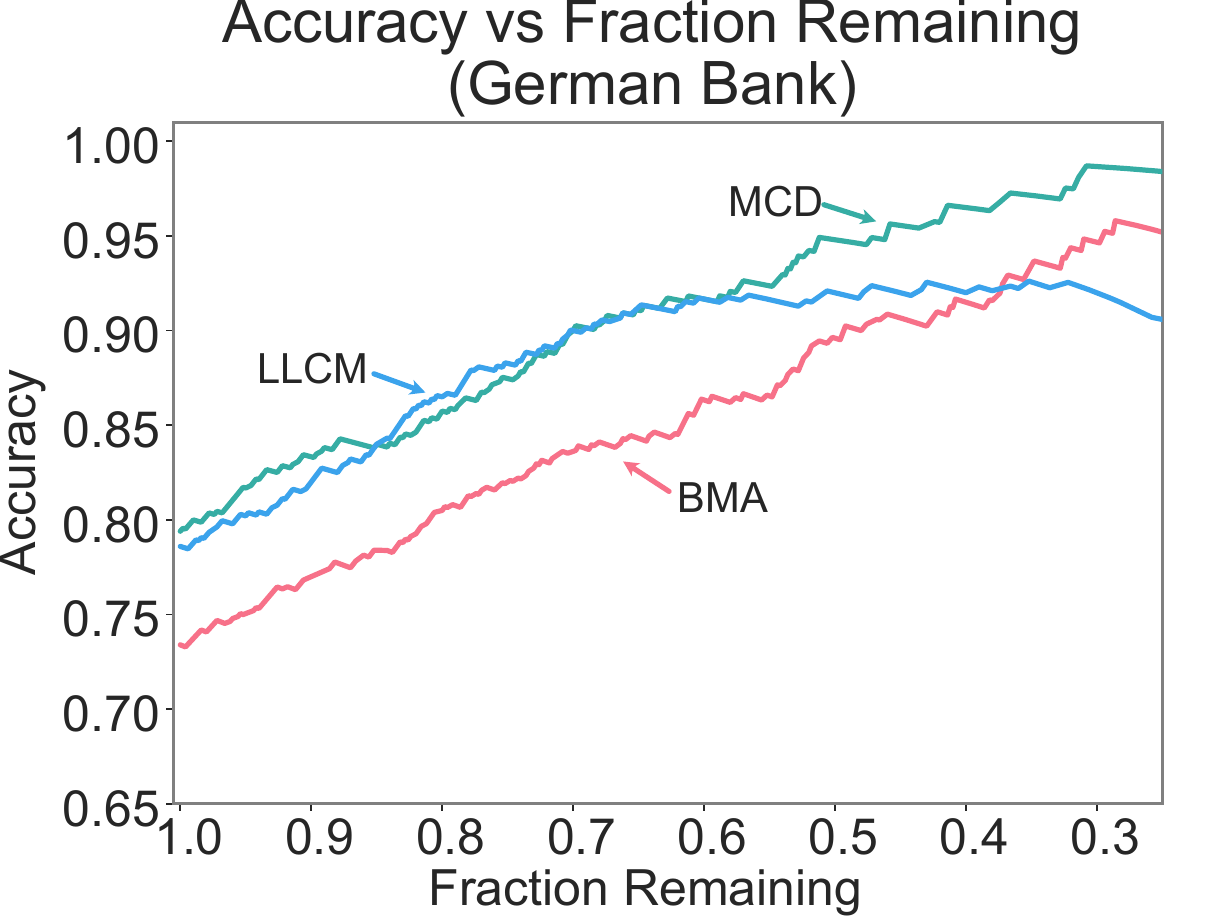}
    \caption{}
    \label{fig:gb_p_scores_accuracy}
    \end{subfigure}
    \hspace{-0.25cm}
    \begin{subfigure}{0.24\textwidth}
        \includegraphics[width=\textwidth]{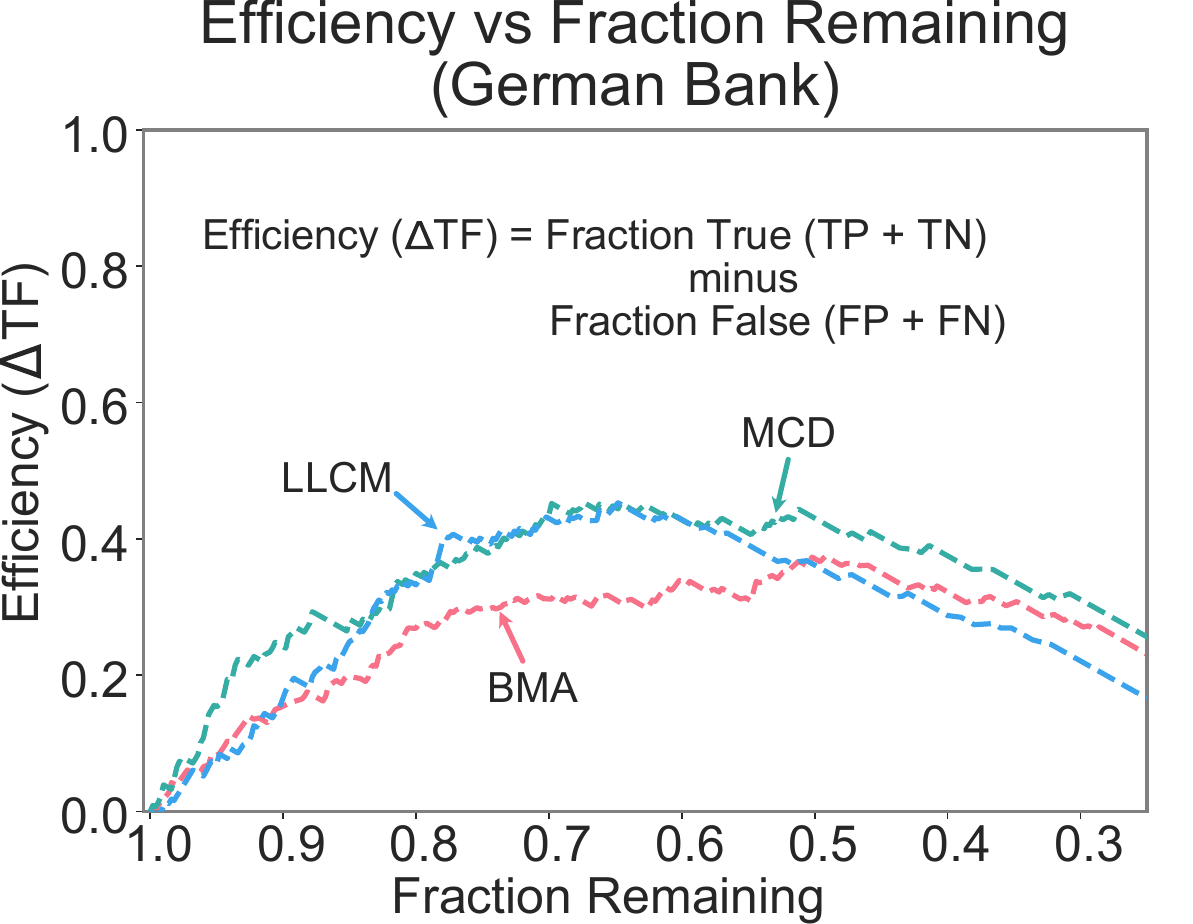}
    \caption{}
    \label{fig:gb_p_scores_efficiency}
    \end{subfigure}
    \vspace{-0.1cm}
    \caption{Uncertain metric plots for the BMA-100, MCD-100 ($-/-/-$), and LLCM-100 ($-/-/-$) networks and the German Bank 2010 dataset using class weights / logit normalization / label smoothing (amount of smoothing, 0.1). These hyperparameters are either applied (denoted by $+$) or omitted (denoted by $-$). Each data point is created by applying increasing thresholds (step size 0.001) of confidence values and re-calculating metrics.}
    \label{fig:gb_p_scores}
\end{figure}

% \FloatBarrier

We next used threshold confidence values to create lists of uncertain samples for the three uncertainty evaluators, BMA-100, MCD-100 ($-/-/-$), and LLCM-100 ($-/-/-$).
To compare the different approaches, we varied threshold values of increasing increments of 0.001, calculated the corresponding accuracy and efficiency and created uncertain metric plots (\autoref{fig:gb_p_scores}).
We opted to plot metrics \textit{vs.}~fraction remaining as this provided a direct interpretation of the fraction of samples to provide an increase in scores.
Both the BMA-100 and MCD-100 provided similar results (albeit, lower accuracy for BMA-100).
The LLCM-100 resulted with near identical accuracies of the MCD-100 up to \textit{ca.} 0.65 of samples remaining, where it plateaus (\autoref{fig:gb_p_scores_accuracy}).
Interestingly, this corresponds to the same point in the uncertain efficiency plot where the maximum difference occurs for correct and incorrect model predictions (\autoref{fig:gb_p_scores_efficiency}).
We observed that both MCD-100 and LLCM-100 behaved near identical for the uncertain efficiency plot, while the BMA-100 resulted in a similar trend with slightly lower performance.

The results from the German Bank 2010 dataset and our LLCM provided comparable performances to BMA-100 and MCD-100; however, using only a single forward-pass of inputs as opposed to 100 sampling of network parameters.
In fact, comparing the use of different hyperparameters resulted with near identical outcomes.

\paragraph{Substrate (depth 2) dataset}
This datasets contains 5 classes of boulders, cobbles, rocks, pebbles/gravel, and sand/mud and is particularly challenging due class imbalance (\textit{i.e.}, 75\% sand/mud).
In \autoref{tab:dataset_s2}, the results for the non-ensemble (ResNet-CNN) and ensemble (BMA-100, MCD-100, and LLCM-100) networks are presented.

As was done for the German Bank 2010 dataset, we initially performed a deterministic ResNet-CNN model using class weighting and/or hyperparameters such as logit normalization and label smoothing (\autoref{tab:dataset_s2}, ResNet-CNN).
This resulted in modest performance for accuracy and calibration metrics.
However, when using class weighting and label smoothing, performance dropped \textit{ca.}~10\% as compared to other configurations.
In this case again, we observed that model calibration decreased when using logit normalization and label smoothing.
The ensemble BMA-100 model performed comparably to the non-ensemble models (\autoref{tab:dataset_s2}, BMA-100 \textit{vs.}~Non-ensemble); however, calibration metrics suggests that the model is poorly calibrated.
We then performed MCD-100 as done previously where we obtained results comparable to non-ensemble models (\autoref{tab:dataset_s2}, MCD-100 \textit{vs.}~Non-ensemble) and a similar set of calibration metrics.
Our LLCM-100 results were near identical with respect to MCD-100 and comparable to BMA-100 (\autoref{tab:dataset_s2}, LLCM \textit{vs.}~MCD-100 \textit{vs.}~BMA-100).

\begin{figure}[!tb]
    \centering
    \begin{subfigure}{0.24\textwidth}
        \includegraphics[width=\textwidth]{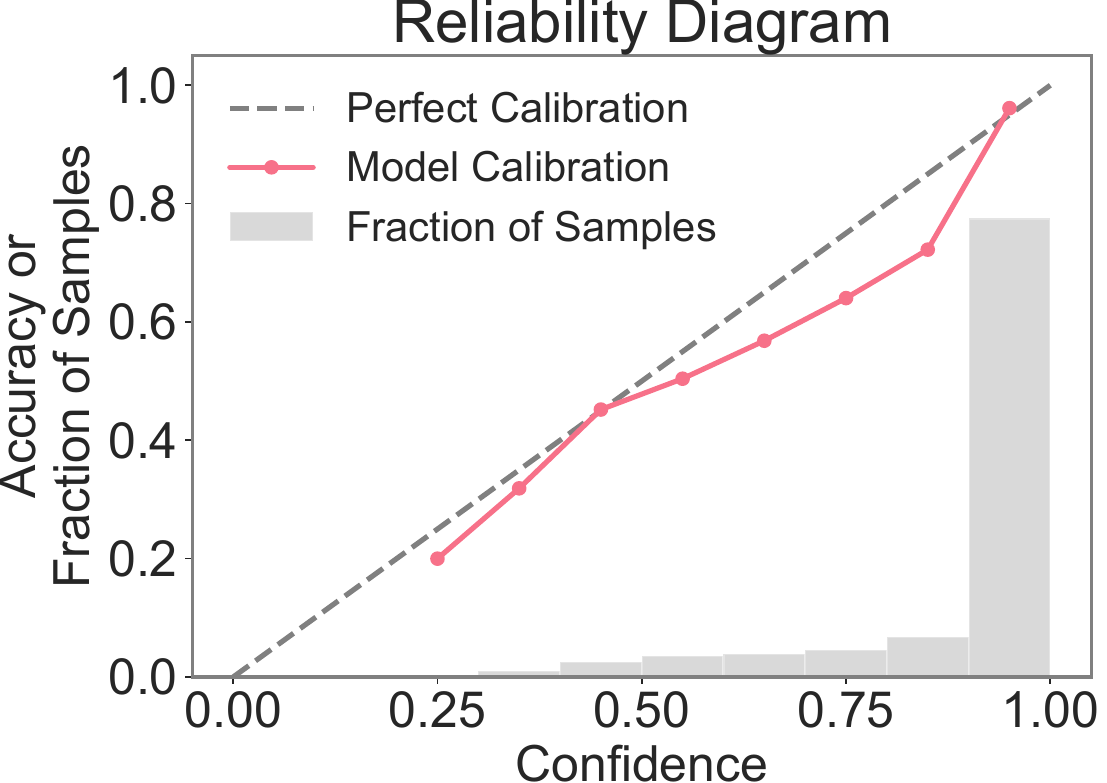}
        \caption{Non-calibrated: $-/-/-$}
        % \label{}
    \end{subfigure}
    \hspace{-0.15cm}
    \vspace{0.25cm}
    \begin{subfigure}{0.24\textwidth}
        \includegraphics[width=\textwidth]{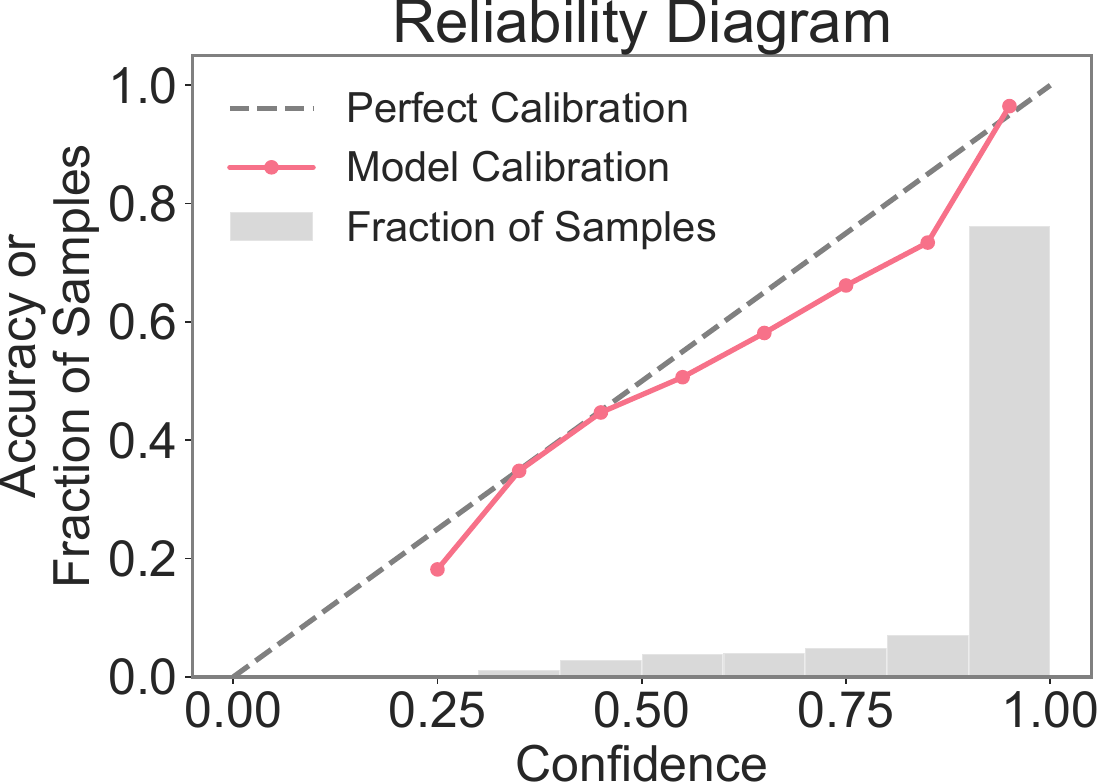}
        \caption{Calibrated: $-/-/-$}
        % \label{}
    \end{subfigure}
    \begin{subfigure}{0.24\textwidth}
        \includegraphics[width=\textwidth]{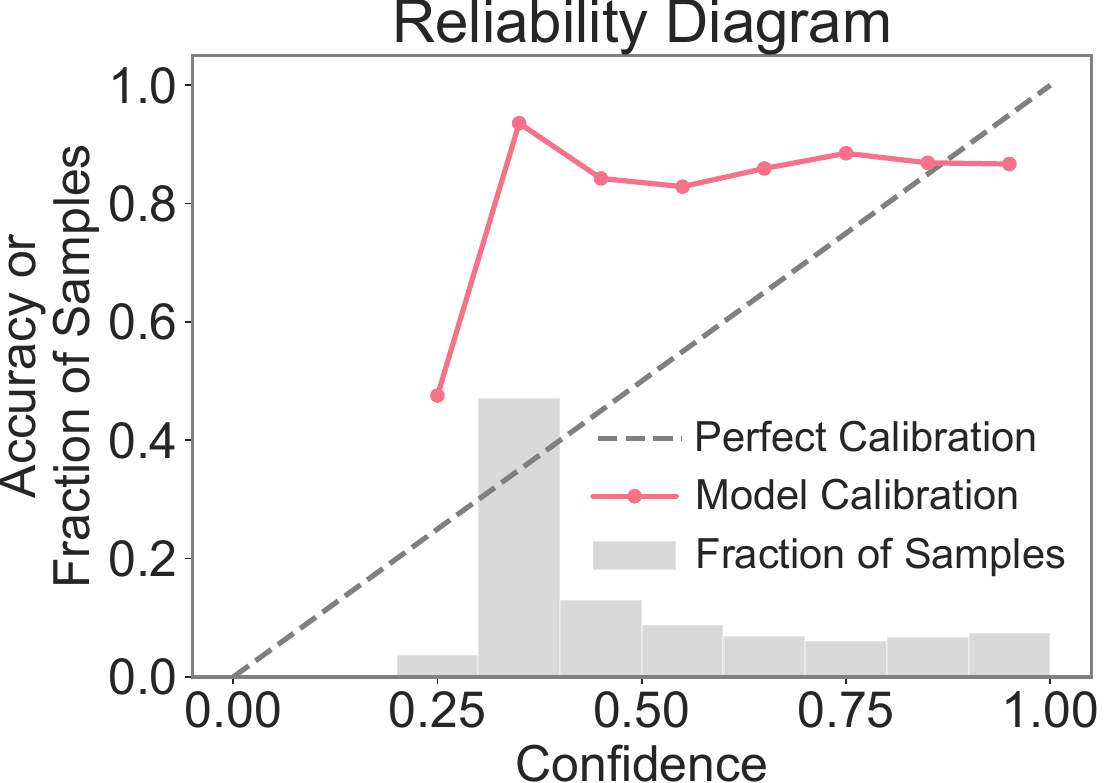}
        \caption{Non-calibrated: $-/+/-$}
        \label{fig:s2_reliability_logit}
    \end{subfigure}
    \hspace{-0.15cm}
    \vspace{0.25cm}
    \begin{subfigure}{0.24\textwidth}
        \includegraphics[width=\textwidth]{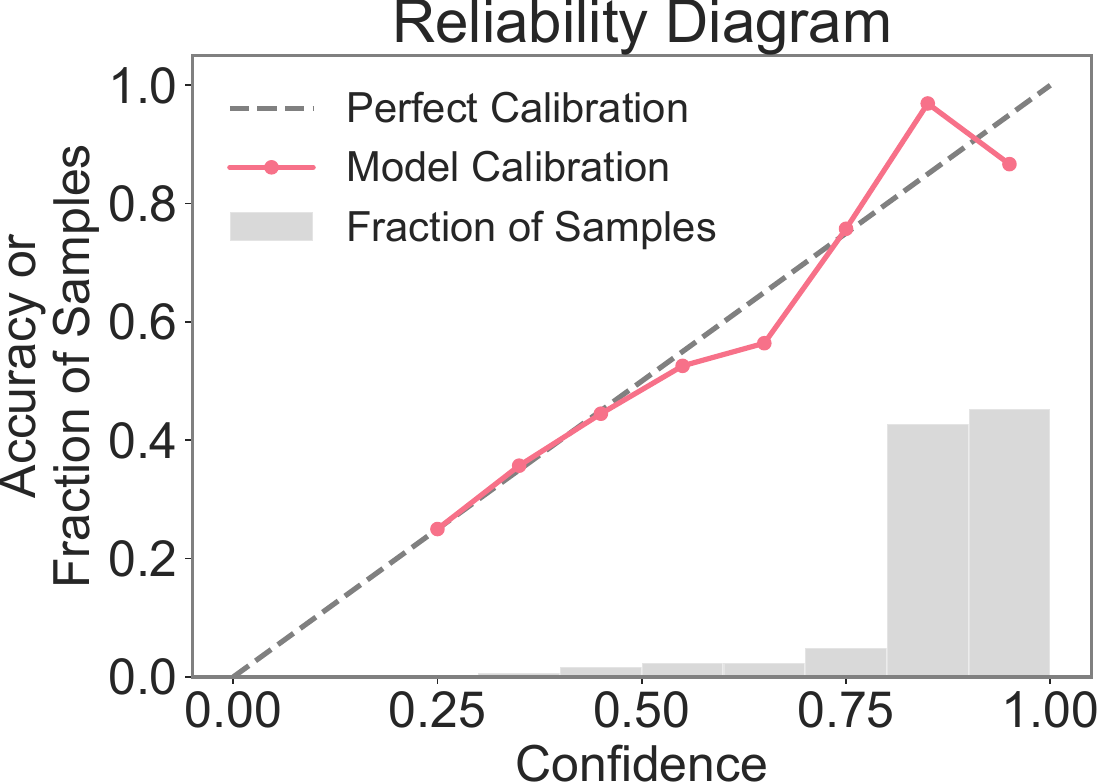}
        \caption{Calibrated: $-/+/-$}
        \label{fig:s2_reliability_calibrated_logit}
    \end{subfigure}
    \begin{subfigure}{0.24\textwidth}
        \includegraphics[width=\textwidth]{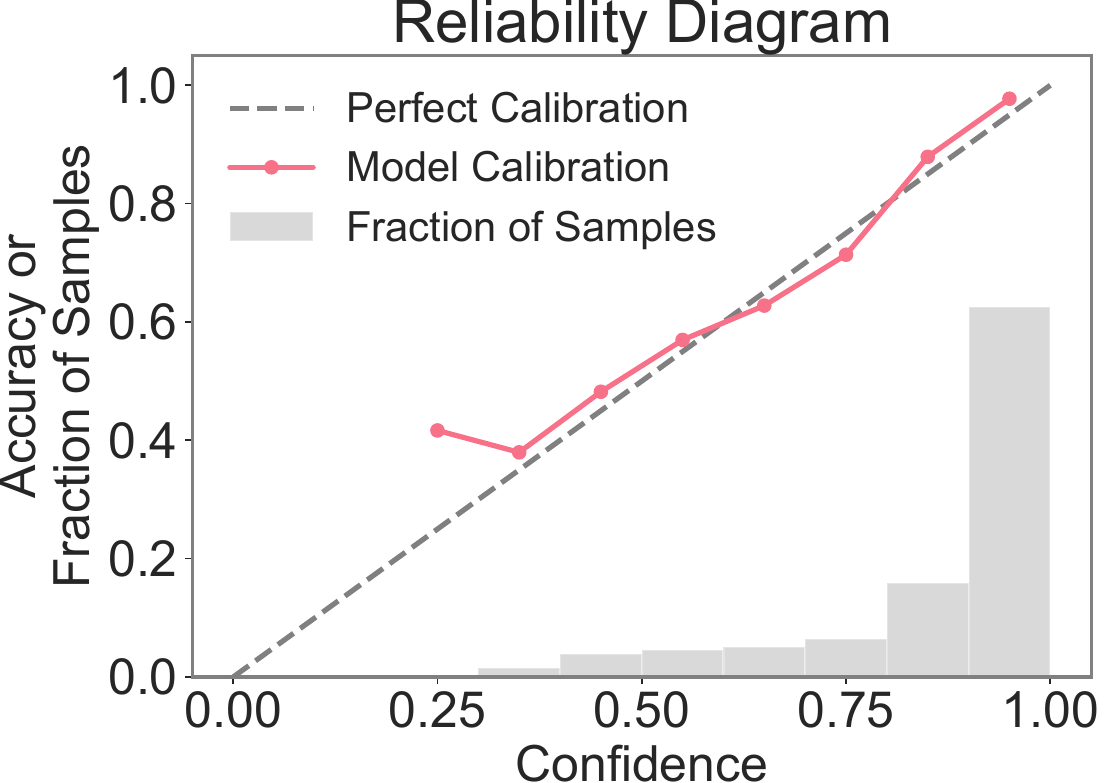}
        \caption{Non-calibrated: $-/-/+$}
        % \label{}
    \end{subfigure}
    \hspace{-0.15cm}
    \vspace{0.25cm}
    \begin{subfigure}{0.24\textwidth}
        \includegraphics[width=\textwidth]{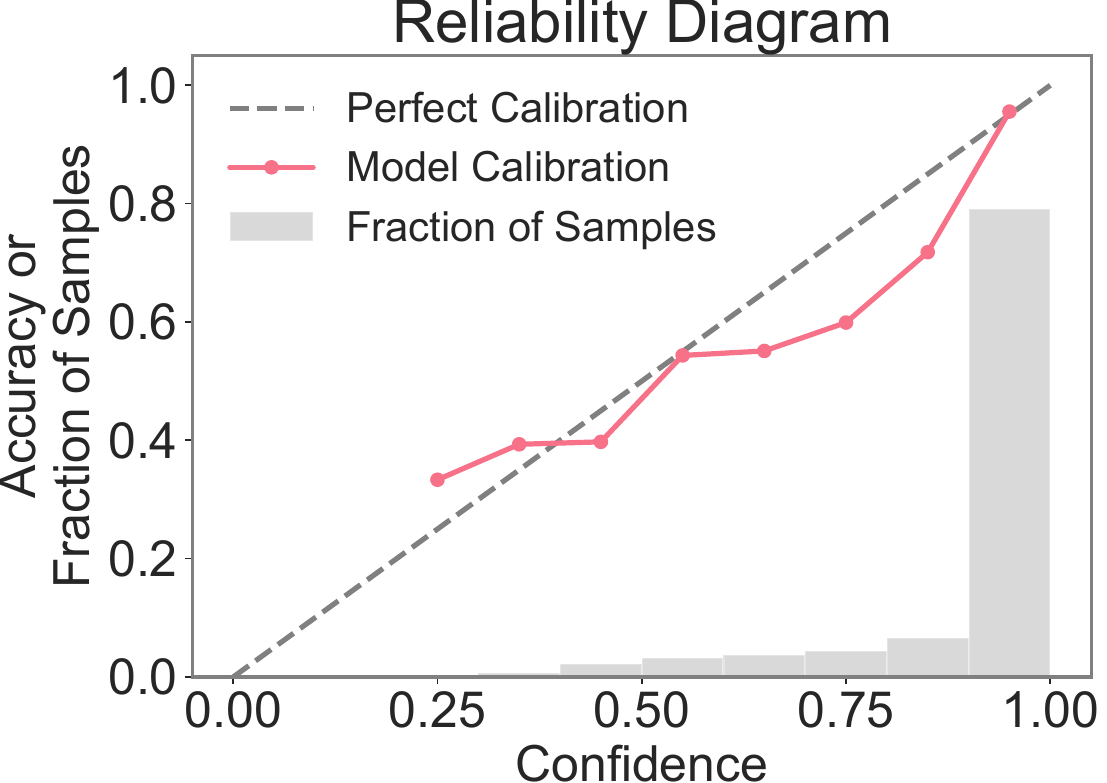}
        \caption{Calibrated: $-/-/+$}
        % \label{}
    \end{subfigure}
    \vspace{-0.75cm}
    \caption{Reliability diagrams for the Substrate (depth 2) dataset using a LLCM-100 network ($p=0.01$). Non-calibrated (left panels) and calibrated (right panels) using class weights / logit normalization / label smoothing (amount of smoothing, 0.1). These hyperparameters are either applied (denoted by $+$) or omitted (denoted by $-$).}
    \label{fig:s2_reliability}
\end{figure}

Reliability diagrams (\autoref{fig:s2_reliability}) for the LLCM-100 non-calibrated and calibrated models using cross-entropy loss with ($-/+/-$) or without ($-/-/-$) logit normalization or label smoothing ($-/-/+$) indicated the largest effect on calibration when using logit normalization (\autoref{fig:s2_reliability_logit}).
As seen with the German Bank 2010 dataset, that can be corrected by applying a re-calibration (\autoref{fig:s2_reliability_calibrated_logit}).

\begin{table}[!tb]
    \centering
    \begin{adjustbox}{width=0.48\textwidth}
    \begin{threeparttable}
        \caption{Model performances using ensemble and non-ensemble models and the Substrate (depth 2) dataset.}
        \begin{tabular}{ l l l l l l }
            \toprule
            % headings
              \textbf{Model}\tnote{1,2}
            & \textbf{Accuracy / F1$\uparrow$}
            & \textbf{NLL$\downarrow$}
            & \textbf{BS$\downarrow$}
            & \textbf{ECE$\downarrow$}
            \\
            \midrule
            \multicolumn{5}{l}{\textbf{Non-ensemble:} ResNet-CNN ($p=0.01$)} \\
            \midrule
            % entry: logs/substrate_depth_2/resnet50p01bt_updated/01; logs/substrate_depth_2/resnet50p01bt_updated_calibrated/01
              $-$/$-$/$-$
            & 0.883 / 0.696
            & 0.094 (0.107)
            & 0.025 (0.029)
            & 0.042 (0.031)
            \\
            % entry: logs/substrate_depth_2/resnet50p01bt_updated/02; logs/substrate_depth_2/resnet50p01bt_updated_calibrated/02
              $-$/$+$/$-$
            & 0.883 / 0.682
            & 0.590 (0.114)
            & 0.206 (0.035)
            & 0.295 (0.039)
            \\
            % entry: logs/substrate_depth_2/resnet50p01bt_updated/03; logs/substrate_depth_2/resnet50p01bt_updated_calibrated/03
              $+$/$+$/$-$
            & 0.825 / 0.641
            & 0.676 (0.179)
            & 0.243 (0.057)
            & 0.284 (0.047)
            \\
            % entry: logs/substrate_depth_2/resnet50p01bt_updated/04; logs/substrate_depth_2/resnet50p01bt_updated_calibrated/04
              $+$/$-$/$-$
            & 0.832 / 0.648
            & 0.142 (0.162)
            & 0.039 (0.044)
            & 0.055 (0.040)
            \\
            % entry: logs/substrate_depth_2/resnet50p01bt_updated/05; logs/substrate_depth_2/resnet50p01bt_updated_calibrated/05
              $+$/$-$/$+$
            & 0.761 / 0.573
            & 0.760 (0.459)
            & 0.287 (0.151)
            & 0.302 (0.234)
            \\
            % entry: logs/substrate_depth_2/resnet50p01bt_updated/06; logs/substrate_depth_2/resnet50p01bt_updated_calibrated/06
              $-$/$-$/$+$
            & 0.882 / 0.691
            & 0.186 (0.100)
            & 0.045 (0.025)
            & 0.038 (0.036)
            \\
            \midrule
            \multicolumn{5}{l}{\textbf{Ensemble: BMA-100} (ResNet-BNN)} \\
            \midrule
            % entry: logs/substrate_depth_2/bnnresnet50bt_updated/02
              BNN
            & 0.830 / 0.531
            & 0.260 
            & 0.082 
            & 0.030 
            \\
            \midrule
            \multicolumn{5}{l}{\textbf{Ensemble: MCD-100} (ResNet-CNN; $p=0.01$)} \\
            \midrule
            % entry: logs/substrate_depth_2/resnet50p01bt_updated/01/dropouts; logs/substrate_depth_2/resnet50p01bt_updated_calibrated/01/dropouts
              $-$/$-$/$-$
            & 0.885 / 0.696
            & 0.100 (0.114)
            & 0.027 (0.031)
            & 0.034 (0.024)
            \\
            % entry: logs/substrate_depth_2/resnet50p01bt_updated/02/dropouts; logs/substrate_depth_2/resnet50p01bt_updated_calibrated/02/dropouts
              $-$/$+$/$-$
            & 0.884 / 0.685
            & 0.624 (0.127)
            & 0.221 (0.039)
            & 0.318 (0.033)
            \\
            % entry: logs/substrate_depth_2/resnet50p01bt_updated/03/dropouts; logs/substrate_depth_2/resnet50p01bt_updated_calibrated/03/dropouts
              $+$/$+$/$-$
            & 0.827 / 0.641
            & 0.723 (0.199)
            & 0.264 (0.063)
            & 0.312 (0.033)
            \\
            % entry: logs/substrate_depth_2/resnet50p01bt_updated/04/dropouts; logs/substrate_depth_2/resnet50p01bt_updated_calibrated/04/dropouts
              $+$/$-$/$-$
            & 0.836 / 0.648
            & 0.153 (0.173)
            & 0.042 (0.047)
            & 0.043 (0.028)
            \\
            % entry: logs/substrate_depth_2/resnet50p01bt_updated/05/dropouts; logs/substrate_depth_2/resnet50p01bt_updated_calibrated/05/dropouts
              $+$/$-$/$+$
            & 0.773 / 0.573
            & 0.768 (0.466)
            & 0.291 (0.154)
            & 0.309 (0.238)
            \\
            % entry: logs/substrate_depth_2/resnet50p01bt_updated/06/dropouts; logs/substrate_depth_2/resnet50p01bt_updated_calibrated/06/dropouts
              $-$/$-$/$+$
            & 0.886 / 0.691
            & 0.191 (0.105)
            & 0.046 (0.026)
            & 0.044 (0.029)
            \\
            \midrule
            \multicolumn{5}{l}{\textbf{Ensemble: LLCM-100} (ResNet-LLCM; $M=100;~p=0.01$)} \\
            \midrule
            % entry: logs/substrate_depth_2/multiresnet50p01bt_updated/01; logs/substrate_depth_2/multiresnet50p01bt_updated_calibrated/01
              $-$/$-$/$-$
            & 0.879 / 0.684
            & 0.106 (0.114)
            & 0.029 (0.031)
            & 0.038 (0.031)
            \\
            % entry: logs/substrate_depth_2/multiresnet50p01bt_updated/02; logs/substrate_depth_2/multiresnet50p01bt_updated_calibrated/02
              $-$/$+$/$-$
            & 0.879 / 0.689
            & 0.741 (0.134)
            & 0.281 (0.028)
            & 0.382 (0.105)
            \\
            % entry: logs/substrate_depth_2/multiresnet50p01bt_updated/03; logs/substrate_depth_2/multiresnet50p01bt_updated_calibrated/03
              $+$/$+$/$-$
            & 0.814 / 0.632
            & 0.528 (0.190)
            & 0.180 (0.048)
            & 0.203 (0.095)
            \\
            % entry: logs/substrate_depth_2/multiresnet50p01bt_updated/04; logs/substrate_depth_2/multiresnet50p01bt_updated_calibrated/04
              $+$/$-$/$-$
            & 0.822 / 0.638
            & 0.158 (0.170)
            & 0.043 (0.046)
            & 0.053 (0.044)
            \\
            % entry: logs/substrate_depth_2/multiresnet50p01bt_updated/05; logs/substrate_depth_2/multiresnet50p01bt_updated_calibrated/05
              $+$/$-$/$+$
            & 0.776 / 0.576
            & 0.761 (0.506)
            & 0.289 (0.173)
            & 0.345 (0.292)
            \\
            % entry: logs/substrate_depth_2/multiresnet50p01bt_updated/06; logs/substrate_depth_2/multiresnet50p01bt_updated_calibrated/06
              $-$/$-$/$+$
            & 0.879 / 0.685
            & 0.187 (0.105)
            & 0.045 (0.026)
            & 0.041 (0.035)
            \\
            \midrule
            \bottomrule
        \end{tabular}
        % add notes using \tnote{#}
        $^1$ BMA: 100 models; MCD: 100 inference samplings; and LLCM: 100 committee members. 
        $^2$ Models defined based on training hyperparameters using class weights / logit normalization / label smoothing (amount of smoothing, 0.1). These hyperparameters are either applied (denoted by $+$) or omitted (denoted by $-$). Values in parentheses are for calibrated models using calculated temperature(s) as previously described. F1-scores reported as macro-averaged.
        \label{tab:dataset_s2}
    \end{threeparttable}
    \end{adjustbox}
\end{table}

% \FloatBarrier

Overall, we observed less of an effect when viewing reliability diagrams of the Substrate (depth 2) dataset compared to the smaller German Bank 2010 dataset.
It would be prudent to not only evaluate calibration metrics, but also to plot reliability diagrams to gain insight relating to under- and/or over-confidence.
In fact, per-class reliability diagrams could be explored for additional details.

% \FloatBarrier

\begin{figure}[!tb]
    \centering
    \begin{subfigure}{0.24\textwidth}
        \includegraphics[width=\textwidth]{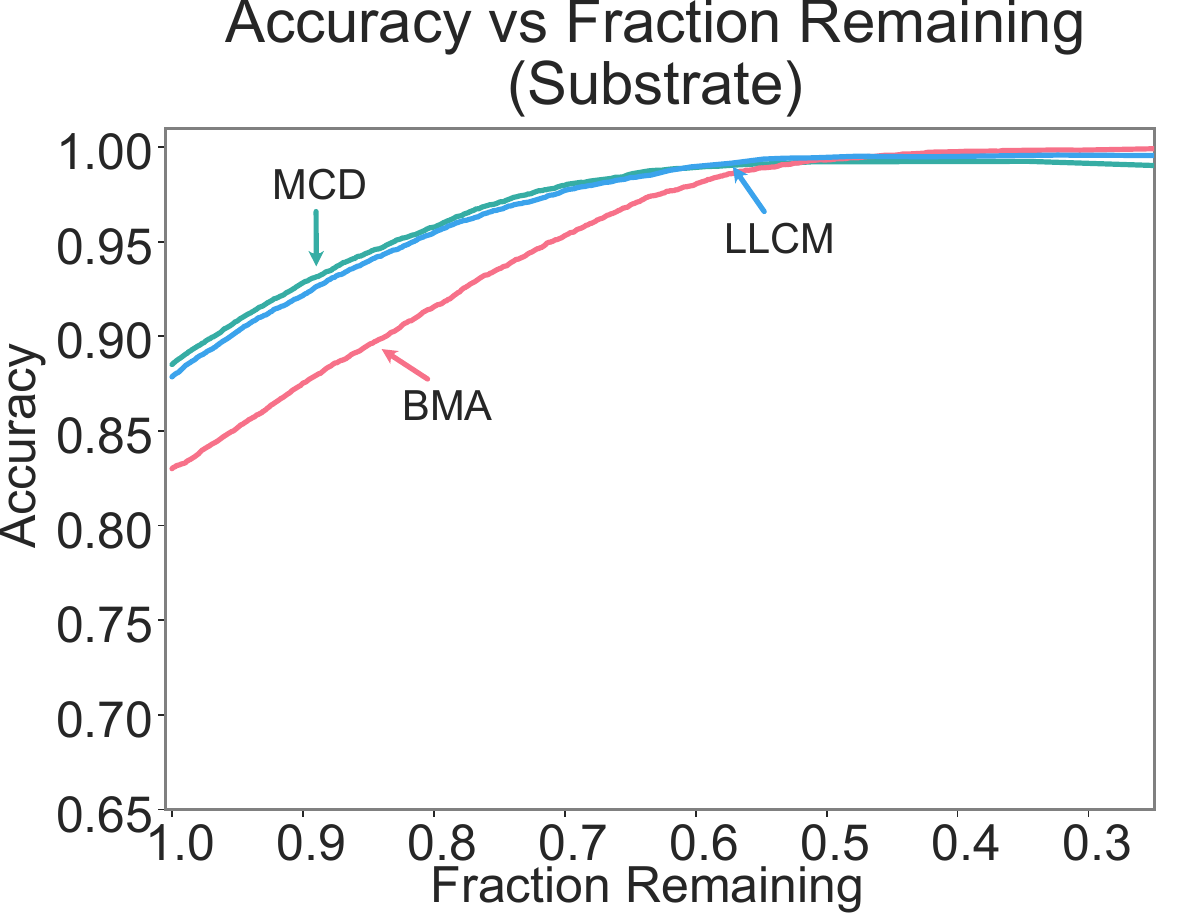}
        \caption{}
        \label{fig:s2_p_scores_accuracy}
    \end{subfigure}
    \hspace{-0.2cm}
    \begin{subfigure}{0.24\textwidth}
        \includegraphics[width=\textwidth]{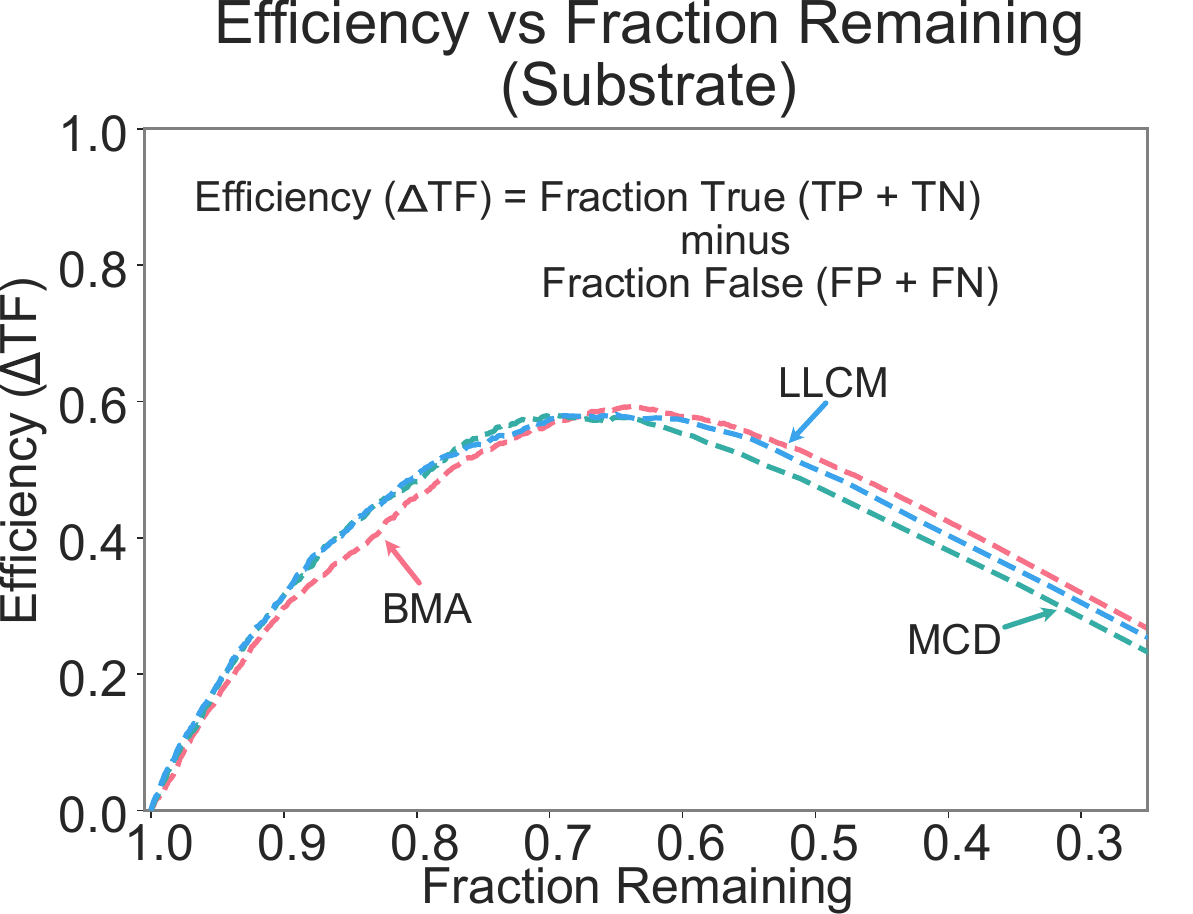}
        \caption{}
    \label{fig:s2_p_scores_efficiency}
    \end{subfigure}
    \vspace{-0.1cm}
    \caption{Uncertain metric plots for the BMA-100, MCD-100 ($-/-/-$), and LLCM-100 ($-/-/-$) networks and the Substrate (depth 2) dataset using class weights / logit normalization / label smoothing (amount of smoothing, 0.1). These hyperparameters are either applied (denoted by $+$) or omitted (denoted by $-$). Each data point is created by applying increasing thresholds (step size 0.001) of confidence values and re-calculating metrics.}
    \label{fig:s2_p_scores}
\end{figure}

We next explored generating lists of uncertain samples for the three uncertainty evaluators, BMA-100, MCD-100 ($-/-/-$), and LLCM-100 ($-/-/-$).
Here again, we varied a threshold of confidence values and calculated the accuracy and efficiency of the remaining samples.
The resulting uncertain metric plots (\autoref{fig:s2_p_scores}) displayed similar trends as seen previously for the German Bank 2010 dataset (\textit{vide supra}).
The overall performance for all uncertainty evaluators were comparable for both the uncertain accuracy and uncertain efficiency plots.
The MCD-100 and LLCM-100 were near identical in regards to the uncertain metric plots, whereas the BMA-100 model was slightly less for the uncertain accuracy plot but essentially the same for the uncertain efficiency plot.
Despite the Substrate (depth 2) dataset being heavily imbalanced, these uncertainty evaluators provided an approach to obtain uncertain samples.

% \FloatBarrier

\paragraph{Inference runtime comparison across uncertainty evaluators}

\autoref{tab:inference_times}~summarizes representative inference runtimes for obtaining per-sample predictions across the benchmark and two benthic imagery datasets. 
The results highlight the computational efficiency of the proposed LLCM relative to BMA and MCD ensembles.
While BMA-100 required 47.6 msec for the Substrate (depth 2) dataset and MCD-100 318.9 msec, the LLCM-100 achieved comparable predictive performance in only 3.0 msec using a single forward pass.
This corresponds to a $> 95$\% reduction in computational time while maintaining comparable performance, which scales consistently across all datasets.
These results supports the central claim that last-layer ensembling provides a scalable, compute-efficient approximation to fully Bayesian (MNIST), last-layer Bayesian (German Bank 2010 and Substrate (depth 2)) or stochastic inference without compromising performance.

\begin{table}[!tb]
    \centering
    \begin{adjustbox}{width=0.48\textwidth}
    \begin{threeparttable}
        \caption{Representative inference runtimes for per-sample (msec) predictions using a 100-member ensemble.}
        \begin{tabular}{ l c c c }
            \toprule
            % headings
            \textbf{Model}\tnote{1}
            & \textbf{MNIST}
            & \textbf{German Bank 2010}
            & \textbf{Substrate (depth 2)}
            \\
            \midrule
            % entry
            BMA-100
            & 2.4 ($60\times$)
            &~~50.4 ($10\times$)
            & 47.6 ($15\times$)
            \\
            % entry
            MCD-100
            & 0.4 ($10\times$)
            & 483.3 ($85\times$)
            & 318.9 ($105\times$)
            \\
            % entry
            LLCM-100
            & 0.04
            & 5.6
            & 3.0
            \\
            \midrule
            \bottomrule
        \end{tabular}
        $^1$ Experiments were conducted on a workstation equipped with an Intel Core i9-13900KF CPU (24 cores, 32 threads), 32 GB RAM, and an NVIDIA GeForce RTX 4090 GPU (24 GB VRAM).
        MNIST: 10000 samples (79 batches); German Bank 2010: 500 samples (4 batches); and Substrate (depth 2): 13719 samples (108 batches). 
        Values in parentheses denote relative inference times with respect to the LLCM-100 on the same dataset.
        \label{tab:inference_times}
    \end{threeparttable}
\end{adjustbox}
\end{table}

% \FloatBarrier

\paragraph{Effects of label smoothing and logit normalization on calibration}

A model is well-calibrated when the predicted confidence reflects the true likelihood of the correct label defined as
$
P(\hat{y} = Y \mid \hat{p}(x) = p) \approx p,
$
where $\hat{y}$ is the predicted class, $Y$ is the true class, $\hat{p}(x)$ is the predicted confidence (\textit{i.e.}, softmax probability).
That is, if the model outputs a confidence of 0.8, it should be correct approximately 80\% of the time. 
Calibration quality is often reported as ECE, which measures the mean deviation between confidence and accuracy across probability bins~(\autoref{par:eval_metrics}).
In reliability diagrams, values above the perfect calibration line indicate under-confident predictions, whereas values below the line correspond to over-confident predictions.

For many examples across all datasets (\autoref{tab:dataset_mnist}--\autoref{tab:dataset_s2}), applying label smoothing or logit normalization increased ECE prior to temperature scaling. 
Label smoothing rescales one-hot targets with attenuated distributions, producing under-confidence, while logit normalization rescales logits by their $\ell_2$-norm (Euclidean norm) prior to the softmax operation, biasing probabilities towards uniform distributions.
As a result, while these methods suppress over-confidence, they impact calibration unless corrected through post-hoc temperature scaling.
In this study, temperature scaling was applied to all models after training, which restored the alignment between confidence and accuracy and reduced the expected calibration error (ECE) to near-baseline levels.
These findings suggest that before introducing confidence-regularization techniques, an initial calibration assessment would be advisable.

In summary, our efforts for the German Bank 2010 and Substrate (depth 2) datasets was focused on establishing an efficient and accessible method to identify uncertain benthic images for review by subject matter experts.
Our long-term goal is to incorporate an active learning framework where data pools of uncertain samples are created and a user-interface is provided for marine scientists to interact and re-evaluate uncertain samples.
While the two commonly used approaches (\textit{i.e.}, BMA and MCD) are options, they can be difficult to configure and require substantial compute time and hardware requirements.
With millions of available benthic images, we require a simple and robust approach to provide prioritized lists of uncertain images.
The LLCM option provides users with a simple and comparable approach, further simplifying existing strategies with reduced compute requirements and complex configurations.

% CONCLUSION
\section{Conclusions}
\label{sect:conclusions}
We compared our LLCM with commonly used Bayesian neural networks and Monte Carlo dropout inference samplings for the difficult benthic datasets, German Bank 2010 and Substrate (depth 2) using available pre-trained benthic models.
We evaluated all networks using accuracy and uncertainty metrics such as NLL, BS, ECE, and in some cases, reliability diagrams.
In addition, we investigated logit normalization and label smoothing as two approaches to mitigate over-confident predictions.
These techniques appeared to degrade model calibration, which can be corrected using a temperature-scaling protocol, where optimal temperature(s) were determined using Bayesian optimization. 

Our results shows that the LLCM is comparable to BMA and MCD methods.
One particular advantage of the LLCM is that only a single forward-pass of inputs is required to obtain per-sample uncertainties.
Whereas, both BMA and MCD require multiple inferences and then average output predictions to obtain per-sample uncertainties.
Given the challenges of labeling benthic imagery, the size of the datasets, and the evolving nature of images in these datasets over time, LLCMs offers an efficient approach for Bayesian approximations.
Prioritized lists of uncertain predictions can be generated and provided to marine scientists to re-evaluate and enhance existing approaches.

% BACKMATTER
\section*{CRediT authorship contribution statement}
\textbf{H. Martin Gillis:} Conceptualization – Ideas; Methodology – Development or design of methodology; creation of models; Software; Writing – original draft;
\textbf{Isaac Xu:} Writing – review \& editing;
\textbf{Benjamin Misiuk:} Writing – review \& editing;
\textbf{Craig J. Brown:} Writing – review \& editing;
\textbf{Thomas Trappenberg:} Supervision; Writing – review \& editing.

\section*{Declaration of competing interest}
The authors declare that they have no known competing financial interests or personal relationships that could have appeared to influence the work reported in this paper.

\section*{Data availability}
The data and pre-trained models used for this study are publicly available from the following sources: BenthicNet:~\cite{Misiuk::2024a,Lowe::2025a}
\& MNIST:~\cite{PyTorch::2025b}.

% REFERENCES
\bibliographystyle{elsarticle-harv}
\bibliography{01-referencesV2}

\end{document}